\documentclass[12pt,letterpaper]{article}

\usepackage{amsmath}
\usepackage{amsfonts}
\usepackage{amssymb}
\usepackage{graphicx}
\usepackage{bm}         
\usepackage{color}
\usepackage[colorlinks=true]{hyperref} 
\usepackage{rotating}
\usepackage{float} 
\usepackage{xcolor}

\newtheorem{defin}{Definition}
\newtheorem{prop}{Proposition}

\newtheorem{ex}{Example}

\begin{document}

\title{\textsc{Level-$k$ Thinking in the Extensive Form}\thanks{We thank the editor and two anonymous reviewers for their helpful comments. We thank Dieter Balkenborg, Russell Cooper, Douglas De Jong, Pietr Evdokimov, Rosemarie Nagel, Thomas Ross, and Aldo Rustichini for sharing their data with us. We also thank participants in the Bay Area Experimental Workshop, 2019, for helpful comments. 
Hang gratefully acknowledges financial support from the Shanghai Rising-Star Program under grant No. 23AQ1403200. Burkhard gratefully acknowledges financial support via ARO Contract W911NF2210282. A prior version of the paper was titled ``Extensive-form level-$k$''.}}

\author{Burkhard C. Schipper\thanks{Department of Economics, University of California, Davis. Email: bcschipper@ucdavis.edu} \and Hang Zhou\thanks{School of Finance, Shanghai University of Finance and Economics. Email: zhouhang@mail.shufe.edu.cn}}

\date{January 18, 2024}
\maketitle

\begin{abstract}
Level-$k$ thinking has been widely applied as a solution concept for games in normal form in behavioral and experimental game theory. We consider level-k thinking in games in extensive form. Player's may learn about levels of opponents' thinking during the play of the game because some information sets may be inconsistent with certain levels. In particular, for any information set reached, a level-$k$ player attaches the maximum level-$\ell$ thinking for $\ell < k$ to her opponents consistent with the information set. We compare our notion of strong level-$k$ thinking with other solution concepts such as level-$k$ thinking in the associated normal form, strong rationalizability, $\Delta$-rationalizability, iterated admissibility, backward rationalizability, backward level-$k$ thinking, and backward induction. We use strong level-$k$ thinking to reanalyze data from some prior experiments in the literature.\newline
\newline
\noindent \textbf{JEL-Classification: } C72, C92, D91. 
\end{abstract}

\thispagestyle{empty}

\newpage

\setcounter{page}{1}

\setlength{\baselineskip}{1.5em}

\section{Introduction}

The core idea of level-k thinking or iterated reasoning about other players is as old as game theory and at the heart of strategic reasoning. Even before the seminal ``Theory of Games and Economic Behavior'' had been published by von Neumann and Morgenstern in 1944, Morgenstern (1928, p. 98) emphasized it in his work on predictions in social situations:\footnote{While the origin of level-$k$ thinking is often traced back to the Beauty Contest described by Keynes in 1936, we note that Morgenstern's work predates Keynes.}

\begin{quote} ``Sherlock Holmes, pursued by his opponent, Moriarty, leaves for Dover. The train stops at a station on the way, and he alights there rather than traveling on to Dover. He has seen Moriarty at the railway station, recognizes that he is very clever, and expects that Moriarty will take a special faster train in order to catch him at Dover. Holmes' anticipation turns out to be correct. But what if Moriarty had been still more clever, had estimated Holmes' mental abilities better and had foreseen his actions accordingly? Then obviously he would have traveled to the intermediate station. Holmes, again, would have had to calculate that, and he himself would have decided to go on to Dover. Whereupon Moriarty would have ``reacted'' differently. Because of so much thinking, they might not have been able to act at all or the intellectually weaker of the two would have surrendered to the other in the Victoria Station, since the whole flight would have become unnecessary. Examples of this kind can be drawn from everywhere.''
\end{quote}

It has been studied in various different forms as sequential best response learning (Cournot, 1838), hierarchies of beliefs (Harsanyi, 1967), iterated admissibility and iterated dominance (Farquharson, 1969, Brams, 1975, Moulin, 1979), rationalizability (Spohn, 1982, Bernheim, 1984, Pearce, 1984), k-level mutual belief in rationality and variants thereof (Tan and Werlang, 1988, Battigalli and Siniscalchi, 2002, Brandenburger, Friedenberg, and Keisler, 2008, Heifetz, Meier, and Schipper, 2019), level-k thinking (Nagel, 1995), and cognitive hierarchies (Stahl and Wilson, 1994, 1995, Camerer, Ho, and Chong, 1994). Latter work on level-k thinking was very much inspired by experiments and has been applied to a wide variety of experimental games (see Crawford, Costa-Gomez, and Iriberri, 2013, for a survey) and seen applications to auctions (Crawford and Iriberri, 2007), mechanism design (Kneeland, 2022), financial market microstructure (Zhou, 2022), and even general equilibrium (Carvajal and Zhou, 2022). We learn from the literature that details matter. In this paper, we focus on details that arise in games in extensive form. We focus on level-$k$ thinking as currently this is a dominant solution concept in experimental game theory thought to capture some notion of bounded rationality.

Although level-$k$ reasoning has been applied to games in extensive form (e.g., Stahl and Haruvy, 2008, Kawagoe and Takizawa, 2005, 2009, 2012, Ho and Su, 2013, Garcia-Pola, Iriberri, and Kovarik, 2020), it is mostly conceived of as a solution concept for games in normal form. Players hold a first-level belief over opponents' behavior (often called level-0 types). A level-1 player best responds to such a first-level belief. A level-2 player best response to his belief that others are level-1 players. A level-3 player best responds to level-2 players etc. The issue is that in games in extensive form, players may learn about the opponents' levels of thinking during the play because some information sets cannot be reached when opponents use certain levels of thinking. Any level of opponents' reasoning that a player learns during play must be below her own level of reasoning. Information about the opponents' level of reasoning is useful for predicting opponents' future play.

In this paper, we present a dynamic version of level-$k$ thinking for games in extensive form titled ``strong level-$k$'' that allows for updating of beliefs over opponents' levels of thinking. We compare strong level-$k$ thinking with level-$k$ thinking in the normal form, strong rationalizability/extensive-form rationalizability, strong $\Delta$-rationalizability, iterated admissibility, backward rationalizability, backward level-k thinking, and backward induction. We focus on the comparison with these solution concepts because all of them can be interpreted as some form of iterative reasoning. We show that for initial full-support beliefs, strong level-$k$ thinking refines level-$k$ thinking outcomes in the normal form. However, while level-$k$ thinking in the normal form refines level-$k$ rationalizability in the normal form, strong level-$k$ thinking does not refine (and is not refined by) $k$-level strong rationalizability (also called $k$-level extensive-form rationalizability). We also show that strong level-$k$ thinking differs from $k$-level strong $\Delta$-rationalizability and from $k$-level iterative admissibility. Finally, we compare it to $k$-level backward rationalizability, backward level-$k$ thinking, and $k$-level backward induction.

Our goal in proposing a notion of strong level-$k$ thinking is not to put up another contender in a horse race of solution concepts that magically predict behavioral data better than any other in games in extensive form. Rather, our hope is that by confronting experimental data with various solution concepts such as strong level-$k$ thinking, level-$k$ thinking in the normal form, and $k$-level strong rationalizability etc., we are able to learn about particular features of human strategic reasoning that are reflected in one solution concept but not in another keeping other features fixed. For instance, by comparing the fit of strong level-$k$ thinking and level-$k$ thinking in the normal form, we can learn about the prevalence of forward induction \emph{given} comparable levels of reasoning and assumptions on level-1 beliefs/level-0 behavior. Or by comparing the fit of strong level-$k$ thinking and $k$-level strong rationalizability, we can learn about the impact of assumptions on level-1 beliefs/level-0 behavior \emph{given} comparable levels of reasoning and the ability to do forward induction. As a first proof of concept, we reanalyze data from versions of the battle-of-the-sexes game with an outside option by Cooper et al. (1993), Balkenborg and Nagel (2016), and Evdokimov and Rustichini (2016).

The closest papers to ours are Stahl and Haruvy (2008), Kawagoe and Takizawa (2005, 2012), Ho and Su (2013), and Lin and Palfrey (2023). Ho and Su (2013) consider repeated play of centipede games where the dynamic aspect concerns learning of levels \emph{between} centipede games. Within each centipede game, each player uses what we would call backward level-$k$ thinking (in analogy to backward rationalizability and backward induction), namely level-$k$ thinking applied to \emph{every} subgame of the centipede game, where the \emph{same} level of thinking is applied to every subgame even if the subgame could not have been reached with such a level of thinking. Ho and Su (2013) allow for updating of levels only \emph{between} runs of the centipede games. We believe that this type of learning about levels from repeated play of the centipede game motivated their terminology of ``dynamic level-$k$ model''.\footnote{Learning of level of reasoning \emph{between} stage games has been studied in various settings by Gil and Prowse (2016), Feng and Wang (2019), and Ho, Park, and Su (2021).} This is different from our approach, which is already dynamic in the one-shot play of a stage game in extensive form. We allow for updating about opponents' levels of thinking \emph{within} the play of a game in extensive form. That is, we do not requiring the same level-$k$ thinking in every subgame and thus we do not force players to potentially ignore information during the play of the stage game that might contradict their belief in their opponents' level of thinking. Updated beliefs about opponents' level of thinking during the play of the stage game is useful for forward inducing opponents' behavior in later subgames of the same stage game and act accordingly. With our notion of strong level-$k$ thinking, players are able to learn about opponents' levels of reasoning from opponents' play throughout the game. Consequently, their own play may vary with what they learned about opponents' levels of reasoning earlier in the game. There is quite some experimental evidence for the assumption that a player's behavior depend on her belief about the levels of reasoning of opponents; see for instance Agranov et al. (2012) and Alaoui, Janezic, and Penta (2020). The assumption is also consistent with the idea that the levels of reasoning displayed by a player might be endogenous (Alaoui and Penta, 2016). Concurrently with Ho and So (2013), backward level-$k$ thinking was also used by Kawagoe and Takizawa (2012) to analyze behavior in experiments on centipede games. The behavior of level-$0$ types in Kawagoe and Takizawa (2012) is assumed to be either uniform  or altruistic. They also analyze a level-$k$ model in the reduced normal-form and a level-$k$ model in agent normal-form. Both Ho and Su (2013) and Kawagoe and Takizawa (2012) confine their analysis to centipede games while we are interested in notions of level-$k$ thinking applicable to any game in extensive form. Earlier, Stahl and Haruvy (2008) presented an experimental study of backward level-$k$ thinking in two-player two-stage games.

After our paper was completed, we learned from Pierpaolo Battigalli about the extension of the cognitive hierarchy model to games in extensive form by Lin and Palfrey (2023).\footnote{Battigalli (2023) shows that the extension of the cognitive hierarchy model to games in extensive form by Lin and Palfrey (2023) is normal-form invariant while Lin and Palfrey (2023) show that it is not \emph{reduced} normal-form invariant.} Different from the ``standard'' model of level-$k$ thinking, the cognitive hierarchy model assumes that a player with level-$k$ thinking has a non-degenerate belief about the levels of opponents. That is, a level-$k$ player assigns non-zero probability to the opponent being level-$\ell$ for all $\ell$ between zero and $k-1$. In their extension of cognitive hierarchy to games in extensive form, Lin and Palfrey (2023) allow players to update their beliefs about opponents' levels throughout the play of the game. However, because players have non-degenerate beliefs, their solution concept does not entail the best-rationalizability principle (Battigalli, 1996) according to which at each information set the opponent is attributed the highest level of rationality consistent with the information set, which is in contrast to our notion of strong level-$k$ thinking. We think  that this should limit the forward induction power of Lin and Palfrey's (2023) extension of cognitive hierarchy to games in extensive form. It is not a defect of their solution concept though. Quite to the contrary, by comparing their extension of cognitive hierarchy to games in extensive form and our notion of strong level-$k$ thinking (which is an extension of the ``standard'' model of level-$k$ thinking), we are able to learn about the prevalence of forward induction in experimental games and its interaction with levels of reasoning. So experimental game theorists are now in the very fortunate situation of having extensions of both the ``standard'' model of level-$k$ thinking and the cognitive hierarchy model to games in extensive form available for analyzing experimental games. Since the non-degenerate beliefs over levels in the cognitive hierarchy model are interpreted in Lin and Palfrey (2023) as ``truncated rational expectations'' and thus as an equilibrium feature, the cognitive hierarchy may be more appropriate in contexts where such equilibrium features could have been already learned while the strong level-$k$ model may be more appropriate in unique or novel contexts. Curiously, Lin and Palfrey (2023) argued that the fact that the ``standard'' level-$k$ model features degenerate beliefs over opponents' levels is a impediment to its extension to games in extensive form. Our work demonstrates that this is not the case. 

The paper is organized as follows: The next section recalls level-$k$ thinking in the normal form and compares it to rationalizability. This sets the stage for Section~\ref{EFLk_section} in which we introduce the definition of strong level-$k$ thinking, compare it to level-$k$ thinking in the normal form, strong rationalizability/extensive-form rationalizability, strong $\Delta$-rationalizability, iterated admissibility, backward rationalizability, backward level-$k$ thinking, and backward induction. In Section~\ref{experiments}, we present a simply reanalysis of data from prior experiments on Battle-of-the-sexes with an outside option.  Proofs are relegated to the appendix.

\section{Level-$k$ Thinking in the Normal Form}

First, we review level-$k$ thinking for games in normal form. This will turn out to be useful when comparing it to strong level-$k$ thinking in the extensive form. We consider finite games in normal form $\langle N, (A_i)_{i \in N}, (u_i)_{i \in N} \rangle$ that consist of a nonempty finite set of players $N = \{1, ..., n\}$ and for each player $i \in N$, a nonempty finite set of actions $A_i$ and a utility function $u_i: A \longrightarrow \mathbb{R}$ with $A := \times_{i \in N} A_i$. As usual, for any player $i \in N$, we denote by $A_{-i} := \times_{j \in N \setminus \{i\}} A_j$ the set of action profiles of player $i$'s opponents.\footnote{We make use of the ``$-i$'' notational convention for any objects index by players. The index ``$-i$'' refers to profiles of player $i$'s opponents' objects.} Denote by $\Delta(A_{-i})$ the set of probability measures on $A_{-i}$. A belief of player $i$ over opponents' actions is denoted by $\beta_i \in \Delta(A_{-i})$.

We say that player $i$'s action $a_i \in A_i$ is \emph{rational} with $\beta_i$ if $a_i$ maximizes player $i$'s expected utility with $\beta_i$. With these definitions in place, we can define the by now standard solution concept of level-$k$ thinking\footnote{Although we use the established terminology, we must admit that we do not know how level-$k$ thinking is related to actual thought processes in the human brain. A better terminology would be ``level-$k$ best response''.} that has been widely applied in experimental game theory.

\begin{defin}[Level-$k$ Thinking in the Normal Form]\label{levelk} Fix a first level belief profile $\beta^1 = (\beta^1_i)_{i \in N}$ with $\beta^1_i \in \Delta(A_{-i})$ for each $i \in N$. Define inductively for each player $i \in N$,
\begin{eqnarray*}
B_i^1(\beta^1) & = & \{\beta^1_i\} \\
L_i^1(\beta^1) & = & \left\{a_i \in A_i : \begin{array}{l} a_i \mbox{ is rational for player } i \mbox{ with belief } \beta^1_i \end{array}\right\} \\
& \vdots & \\
B_i^k(\beta^1) & = & \left\{\beta_i \in \Delta(A_{-i}) : \beta_i(L_{-i}^{k - 1}(\beta^1)) = 1 \right\} \\
L_i^k(\beta^1) & = & \left\{a_i \in A_i : \begin{array}{l} \mbox{There exists } \beta_i \in B_i^k(\beta^1) \mbox{ such that } a_i \\ \mbox{is rational for player } i
\mbox{ with belief } \beta_i. \end{array} \right\}.
\end{eqnarray*} For $k \geq 1$, we call $L_i^k(\beta^1)$ the set of (normal-form) level-$k$ thinking actions of player $i$ anchored by the profile of first-level beliefs $\beta^1$.
\end{defin}

Several remarks are in order: First, the first-level belief $\beta^1$ is often interpreted as the behavior of level-$0$ players. Since it is also assumed that there are no actual level-$0$ players, we model level-$0$ players more appropriately as what they are, namely just beliefs of level-$1$ players.

Second, level-$k$ thinking is not one solution concept but a collection of solution concepts, one for each first-level belief $\beta^1$ / level-$0$ behavior and each level $k$. In applications, the first-level belief is often fixed to the uniform measure (e.g., Nagel, 1995) although in some applications other distributions are more natural (e.g., Arad and Rubinstein, 2012). These assumptions seem to reflect either unpredictable behavior (i.e., in the spirit of the principle of insufficient reason) or non-strategic level-$0$ behavior.

Third, when iterated best responses are unique, i.e., the sets $L_i^k(\beta^1)$, $k \geq 1$, are singleton, then also $B_i^{k + 1}(\beta^1)$, $k \geq 1$, are singleton. In such a case, it does not matter whether or not we allow for correlated beliefs or, alteratively, would restrict to independent beliefs. Applications focus sometimes on games in which best responses are unique or a best response is selected by some additional ad hoc assumption. In our general approach, we allow for multiple best responses at any level $k$ and allow for any $k+1$-level belief over those $k$-level best responses. 

Fourth, to emphasize that the entire procedure with both beliefs and actions at higher levels are anchored by the first-level belief $\beta^1$, we explicitly write $B_i^k(\beta^1)$ and $L_i^k(\beta^1)$.

\subsection{Level-$k$ Thinking versus Rationalizability\label{NF_levelk_rat}}

The idea of using iterated reasoning about opponents play in a solution concept featured already in solution concepts developed earlier in game theory. In particular, rationalizability introduced by Spohn (1982), Bernheim (1984), and Pearce (1984) is defined inductively using player's beliefs about rational actions of opponents.\footnote{Like most of the literature, we focus on correlated rationalizability; see Brandenburger and Dekel, 1987, and Tan and Werlang, 1988, for more details on rationalizability.}

\begin{defin}[Rationalizability]\label{rationalizable1} Define inductively for each player $i \in N$,
\begin{eqnarray*}
B_i^1 & = & \Delta(A_{-i}) \\
R_i^1 & = & \left\{a_i \in A_i : \begin{array}{l} \mbox{There exists } \beta_i \in B_i^1 \mbox{ such that } a_i \\ \mbox{is rational for player } i \mbox{ with belief } \beta_i. \end{array} \right\} \\
& \vdots & \\
B_i^k & = & \left\{\beta_i \in \Delta(A_{-i}) : \beta_i(R_{-i}^{k - 1}) = 1 \right\} \\
R_i^k & = & \left\{a_i \in A_i : \begin{array}{l} \mbox{There exists } \beta_i \in B_i^k \mbox{ such that } a_i \\ \mbox{is rational for player } i \mbox{ with belief } \beta_i. \end{array} \right\}
\end{eqnarray*} For any $k \geq 1$, we call $R_i^k$ the set of $k$-level rationalizable actions of player $i$. The set of player $i$'s rationalizable actions is
\begin{eqnarray*} R_i^{\infty} & = & \bigcap_{k = 1}^{\infty} R_i^k
\end{eqnarray*}
\end{defin}

It is well-known that rationalizability is strategy-equivalent to iterated elimination of strictly dominated actions (see Pearce, 1984). In fact, for any level $k \geq 1$, level-$k$ rationalizability (i.e., rationalizability up to level $k$) is strategy-equivalent to $k$-iterated elimination of strictly dominated actions. An action is strictly dominated if there exists a possibly mixed action that yields a strictly higher expected utility no matter what opponents play. Thus, rationalizability does not only provide a prediction in the limit when $k$ goes to infinity, but also for every finite level $k$. This has been previously used in experiments to partially identify levels of beliefs (for an approach along these lines see Li and Schipper, 2020). In our context, it is now natural to ask about how behavior implied by level-$k$ thinking is related to level-$k$ rationalizable actions in games in normal form. For any first-order belief (i.e., any level-$0$ behavior), the level-$k$ behavior is $k$-level rationalizable. That is, level-$k$ thinking implies $k$-level rationalizability.

\begin{prop}\label{relationship} For any finite game in normal form, $\langle N, (A_i)_{i \in I}, (u_i)_{i \in I} \rangle$, any profile of first-level beliefs, $\beta^1$, and level $k \geq 1$, $L^{k}(\beta^1) \subseteq R^{k}$.
\end{prop}

While obvious at the first level, the proof by induction in the appendix reveals that at any level $k$, the belief that a player with level-$k$ thinking entertains about opponents' play is a belief that $k$-level rationalizes her action. That is, the inclusion holds not only for actions but also beliefs.

The most obvious difference between rationalizability and level-$k$ thinking is that rationalizability does not fix first-level beliefs. Initially, it allows for \emph{any} first-level beliefs over opponents' actions. This is useful when there is no ``natural'' first-level belief and when it is reasonable to assume that players could entertain any initial belief. Based on Proposition~\ref{relationship}, one may be tempted to claim that level-$k$ reasoning yields sharper predictions than rationalizability. Yet, level-$k$ thinking does not explain the first-level beliefs. That is, it does not provide a theory of first-level beliefs or level-0 behavior. Rather, \emph{given} a first-level belief of players / assumption of level-$0$ behavior of the analyst that, while extremely useful, is necessarily ad hoc because it is outside the model of level-$k$ thinking, it yields a sharper prediction than if the analyst considers any first-level belief as in level-$k$ rationalizability.

In order to emphasize that (1) first-level beliefs of level-$k$ thinking yield strong refinement power, and (2), that level-$k$ thinking differs from rationalizability just in the first-level belief, we state the following weaker converse to Proposition~\ref{relationship}. For every rationalizable action there exists a first-level belief with which the action is rationalizable with level-$1$ thinking. Note that this implies that we can find first-level beliefs that justify any arbitrarily high $k$-level rationalizable action with level-$1$ thinking. Thus, ex ante tying our hands to a particular first-level belief/assumption of level-0 behavior is crucial in experiments if level-$k$ thinking is to have predictive power beyond rationalizability and the identification of levels is to be meaningful. 

\begin{prop}\label{converse} For any finite game in normal form, $\langle N, (A_i)_{i \in I}, (u_i)_{i \in I} \rangle$, every player $i \in N$, and every $a_i \in R_i^{\infty}$, there exists a first-level belief $\beta^1 \in \Delta(A_{-i})$ such that $a_i \in L_i^1(\beta^1)$.
\end{prop}

While the observation might be obvious to some, we present a short proof in the appendix. 

It is also worth emphasizing another difference between level-$k$ thinking and rationalizability: Latter is a reduction procedure on beliefs that implies a reduction procedure on actions while for former this is not necessarily the case. This is demonstrated in the following simple example:

\begin{ex}\label{cycle} Consider the following the following variant of the matching pennies game:
\begin{eqnarray*}\begin{array}{c} \\ U \\ D \end{array} \begin{array}{c} \begin{array}{ccc} L & & R \end{array} \\
\left(\begin{array}{cc} 2, -1 & -1, 2 \\ -1, 1 & 1, -1 \end{array}\right)\end{array}
\end{eqnarray*} Apply level-$k$ thinking anchored with a uniform first-level belief for each player. Then the sets of level-$k$ thinking actions are derived subsequently by
$$\begin{array}{|c|c|c|} \hline & \mbox{ Row player } & \mbox{ Column player} \\ \hline
L_i^1(\beta^1) & \{U\} & \{R\} \\
L_i^2(\beta^1) & \{D\} & \{R\} \\
L_i^3(\beta^1) & \{D\} & \{L\} \\
L_i^4(\beta^1) & \{U\} & \{L\} \\
L_i^5(\beta^1) & \{U\} & \{R\} \\
\vdots & \vdots & \vdots \\ \hline
\end{array}$$ We note that level-$k$ thinking results in a choice cycle. So clearly, it is not the case that the set of level-$k$ actions are refined level by level.
\end{ex}

The fact that level-$k$ thinking is not necessarily a reduction procedure is certainly not desirable from both a strategic point of view and the epistemic point of view. Yet, it matches curiously with the reasoning reflected in the quote from Morgenstern (1928) that we stated in the introduction.

\section{Level-$k$ Thinking in the Extensive Form\label{EFLk_section}}

Consider a finite game in extensive form with possibly imperfect information, perfect recall, finite horizon, and possibly simultaneous moves $\langle N, H, P, (\mathcal{I}_i)_{i \in N}, (u_i)_{i \in N} \rangle$ defined by
\begin{itemize}
\item A nonempty finite set of players $N$.
\item A set $H$ of finite sequences of action profiles (i.e., histories) such that
\begin{itemize}
\item $\emptyset \in H$,
\item If history $(a^m)_{m = 1, ..., M} \in H$ and $M' < M$, then also the subhistory $(a^m)_{m = 1, ..., M'} \in H$.
\end{itemize}
The set of terminal histories are histories with no successors. We denote them as usual by $Z$.

\item A player function $P: H \setminus Z \longrightarrow 2^{N \cup \{c\}} \setminus \{\emptyset\}$ that assigns to each nonterminal history $h \in H \setminus Z$ a nonempty subset of players $P(h) \subseteq N \cup \{c\}$ that may include nature, $c$. That is, players are allowed to move simultaneously. Moreover, nature is allowed to move any time and even simultaneously with other players.

With this notation, let $A_i(h)$ be the nonempty set of actions of player $i \in P(h)$ at the non-terminal history $h \in H \setminus Z$. Moreover, we let $a \in \times_{i \in P(h)} A_i(h)$ be the action profile of players moving at history $h$. That is, $h \in H \setminus Z$ and $a \in \times_{i \in P(h)} A_i(h)$, then $(h, a) \in H$.

\item For each player $i \in N$, a partition $\mathcal{I}_i$ of non-terminal histories in $H_i = \{h \in H \setminus Z : i \in P(h)\}$ at which he moves. Elements of $\mathcal{I}_i$ are called information sets of player $i$. As usual, we require that for any information set $I_i \in \mathcal{I}_i$, $A_i(h) = A_i(h')$ for any $h, h' \in I_i$. Thus, we can simply denote by $A_i(I_i)$ the set of player $i$'s actions at information set $I_i$. Each player's information sets are required to satisfy perfect recall.

\item For each player $i \in N$, there is a von Neumann-Morgenstern utility function over lotteries of terminal histories. We denote by $u_i: Z \longrightarrow \mathbb{R}$ player $i$'s Bernoulli utility index.

\end{itemize}
See for instance Osborne and Rubinstein (1995, Chapters 6.3.2 and 11.1.2) for further details on games in extensive form including perfect recall. We allow for simultaneous moves of players. We also allow for imperfect information and moves of nature at any time during the game including simultaneously with other players. We do not have to assume a prior probability measure over moves of nature but such an assumption can be added whenever it is required.

For any player $i \in N$, a (pure) strategy of player $i$ assigns to each of her information sets an action available at that information set. Formally, a strategy $s_i$ is a function $s_i: \mathcal{I}_i \longrightarrow \bigcup_{I_i \in \mathcal{I}_i} A_i(I_i)$ such that $s_i(I_i) \in A_i(I_i)$ for all $I_i \in \mathcal{I}_i$. Let $S_i$ denote the set of player $i$'s strategies. Let $N_c : = N \cup \{c\}$. We treat nature like a player with information sets that are singletons who is indifference among all terminal histories. Define $S := \times_{i \in N_c} S_i$ if nature moves in the game. Otherwise, $S: = \times_{i \in N} S_i$. Similarly, for any $i \in N$, $S_{-i} := \times_{j \in N_c \setminus \{i\}} S_j$ if natures moves in the game. Otherwise, $S_{-i} := \times_{j \in N \setminus \{i\}} S_j$.

For every player $i \in N$, we say a strategy $s_i \in S_i$ reaches an information set if there exists a profile of opponents' strategies $s_{-i}$ such that $(s_i, s_{-i})$ reaches this information set. Similarly, a profile of opponents' strategies $s_{-i}$ reaches an information set if there exists a strategy $s_i$ of player $i$ such that $(s_i, s_{-i})$ reaches this information set.

A belief system of player $i \in N$, $$\bar{\beta}_i := \left(\bar{\beta}_i(I_i)\right)_{I_i \in \mathcal{I}_i} \in \prod_{I_i \in \mathcal{I}_i} \Delta(S_{-i}),$$ is a profile of beliefs, a belief $\bar{\beta}_i(I_i) \in \Delta(S_{-i})$ about other players' strategies for each information set $I_i \in \mathcal{I}_i$, with the following properties:
\begin{itemize}
	\item $\bar{\beta}_i(I_i)$ reaches $I_i$, i.e., $\bar{\beta}_i(I_i)$ assigns probability 1 to the set of strategy profiles of the other players that reaches $I_i$.
	
	\item If information set $I_i$ precedes information set $I_i'$, then $\bar{\beta}_i(I_i')$ is derived from $\bar{\beta}_i(I_i)$ by conditioning whenever possible.
\end{itemize}

Denote by $\bar{B}_i$ the set of all belief systems of player $i \in N$.

For a player $i$ and an information set $I_i$, a strategy $s_i'$ is a $I_i$-replacement of strategy $s_i$ if $s_i'$ agrees with $s_i$ on all information sets strictly preceding $I_i$.

With a belief system $\bar{\beta}_i$, strategy $s_i$ is rational for player $i$ at the information set $I_i$ if $s_i$ does not reach $I_i$ or if $s_i$ does reach $I_i$ but there does not exist an $I_i$-replacement of $s_i$ which yields a strictly higher expected utility given $\bar{\beta}_i(I_i)$ on $S_{-i}$.

\begin{defin}[Strong Level-$k$ Thinking]\label{EFlevelk} Given a belief system of first-level beliefs, $\bar{\beta}^1 = (\bar{\beta}_i^1)_{i \in N}$ with $\bar{\beta}_i^1 \in B_i$, define inductively for all $i \in N$,
\begin{eqnarray*}
\bar{B}_i^1(\bar{\beta}^1) & = & \{\bar{\beta}_i^1\} \\
\bar{L}_i^1(\bar{\beta}^1) & = & \left\{ s_i \in S_i : \begin{array}{l} \mbox{For every information set } I_i \in \mathcal{I}_i, \\ s_i \mbox{ is rational at } I_i \mbox{ with respect to } \bar{\beta}_i^1. \end{array} \right\} \\
& \vdots & \\
\bar{B}_i^k(\bar{\beta}^1) & = & \left\{ \bar{\beta}_i \in \bar{B}_i : \begin{array}{l} \mbox{For every information set } I_i \in \mathcal{I}_i,
\mbox{if there exists } \ell \\ \mbox{with } 1 \leq \ell < k
\mbox{ for which there exists } s_{-i} \in \bar{L}_{-i}^{\ell}(\bar{\beta}^1) \\
\mbox{such that } s_{-i} \mbox{ reaches } I_i, \mbox{ let } \bar{\ell} \mbox{ be
the largest such } \ell. \\ \mbox{Then } \bar{\beta}_i(I_i)
\mbox{ assigns probability } 1 \mbox{ to } \bar{L}_{-i}^{\bar{\ell}}(\bar{\beta}^1). \mbox{ Otherwise, } \\ \mbox{if there is no such } \ell, \mbox{ then let } \bar{\beta}_i(I_i) = \bar{\beta}^1_i(I_i).
\end{array}\right\} \\
\bar{L}_i^k(\bar{\beta}^1) & = & \left\{ s_i \in S_i : \begin{array}{l} \mbox{There exists } \bar{\beta}_i \in \bar{B}_i^k(\bar{\beta}^1) \mbox{ with which for every } \\\mbox{information set } I_i \in \mathcal{I}_i,  s_i \mbox{ is rational at } I_i. \end{array} \right\}
\end{eqnarray*}	 For any level $k$, we call $\bar{L}^k_i(\bar{\beta}_i)$ player $i$'s set of strong level-$k$ thinking strategies.
\end{defin}

Several remarks are in order: First, a player's assumption of opponents' level-0 behavior is modeled as level-1 belief. However, we like to emphasize that since a player may move at various information sets of the game, she forms possibly different beliefs about opponents' behavior at each of her information sets. According to the notion of belief system, each such a belief must be consistent with having reached this information set and derived by conditioning whenever possible. That is, different from level-$k$ thinking in the normal form, a player not only forms beliefs about opponents' behavior before playing the game but also at each of her information sets throughout the game. The assumption of level-0 behavior of opponents is now a collection of assumptions, one belief at each of her information sets.

Second, strong level-$k$ thinking features the best rationalizability principle (Battigalli, 1996). At each of her information sets, a player following strong level-$k$ thinking assigns the highest possible level $\ell$-thinking with $\ell < k$ to opponents that is consistent with reaching the information set. Intuitively, a player does not easily label opponents' behavior as a ``mistake''. Rather a player tries to make sense of opponents' behavior as much as it possible within her own (limited) thinking. Such an approach makes quite some sense when allowing for learning from opponents' play. If opponents' play is judged easily as a mistake, then there is not much to learn from. Note that the best rationalizability principle does not rule out that a sophisticated player mimics a strategy of a player with a lower level of reasoning. For instance, if a strong level-$k$ player learns at some information set that his opponent is just strong level-$\ell$ with $\ell < k - 1$, then such player may mimic the strategy of a player who is just strong level $\ell + 1$ even though she is capably of higher order reasoning and thus more sophisticated behavior.

Third, a player's belief is updated to lower levels of opponents' thinking along the path of play whenever information sets are reached that cannot be reached with strategies of opponents' featuring higher levels of thinking. Of course, this depends on the player's own level of thinking since one defining feature of level-$k$ thinking is a form of naivet\'{e}, namely that a player with level $k$ must believe opponents' feature levels of thinking strictly lower than level $k$.

Fourth, the phrasing of the definition of $\bar{B}_i^k(\bar{\beta}^1)$ appears somewhat awkward at the first glance, since it involves an ``if ... then ... Otherwise ...'' clause.  Partially, this captures the best rationalizability principle above (i.e., the ``then'' statement). Yet, the ``if'' and ``Otherwise'' clauses emphasize that an information set may not be reachable with any strong level-$\ell$ thinking strategy, for $\ell < k$. What shall a player believe in such a case? We assume that in such a case, the player resorts nevertheless to his first-level belief. Such an assumption is innocuous if the first-level belief is a full support belief like for instance uniform belief as often assumed in the literature on level-$k$ thinking.

Finally, we need to comment on the terminology. The name ``strong level-$k$'' is chosen in analogy to ``strong rationalizability'', which is also called more descriptively ``extensive-form rationalizability'', a notion of rationalizability for games in extensive form that we will introduce later in the text. It is a somewhat unfortunate terminology because ``strong'' is a rather generic adjective. However, it refers to the fact that it features a notion of ``strong'' belief that is, certainty of an event at all information sets consistent with the event. 

In the following subsections, we explore strong level-$k$ thinking by comparing it to various other solution concepts.

\subsection{Normal-Form versus Strong Level-$k$ Thinking}

The key difference between level-$k$ thinking in the normal form and strong level-$k$ thinking is that a player with strong level-$k$ thinking can update beliefs about opponents' level-$\ell$ thinking, for $\ell < k$, conditional on information sets reached. At each information set reached, player $i$ with level-$k$ thinking attributes the highest level-$\ell$ thinking, $\ell < k$, to opponents consistent with the information set. It embodies the best rationalization principle (Battigalli, 1996). To what extent does the best rationalizability principle matter? First, we show that, if $\bar{\beta}^1$ is a profile of full-support belief systems, then strong level-$1$ strategies are equivalent to normal-form level-$1$ strategies.

\begin{prop}\label{EFL1_NFL1} Consider any finite game in extensive form with perfect recall. Let $\bar{\beta}^1$ be a profile of full-support belief systems and $\beta^1$ a profile of full-support beliefs in the associated normal form consistent with $\bar{\beta}^1$. Then for any player $i \in N$, $\bar{L}_i^1(\bar{\beta}^1) = L_i^1(\beta^1)$.
\end{prop}

The observation implies in particular that if beliefs are uniform both in the game in extensive form and the associated normal form (as frequently assumed in experimental work), then level-$1$ thinking strategies coincide.
\begin{sidewaystable}[!p]\caption{Solutions to various games\label{solutions_table}}
\tiny
\centering
\begin{tabular}{||c||c|c||c|c||c|c||c|c||c|c||} \hline \hline Reny (1992) & \multicolumn{2}{c||}{Strong level-$k$} & \multicolumn{2}{c||}{Level-$k$ in the normal form} & \multicolumn{2}{c||}{$k$-level strong} & \multicolumn{2}{c||}{Backward} & \multicolumn{2}{c||}{$k$-level backward} \\
game & \multicolumn{2}{c||}{(uniform)} & \multicolumn{2}{c||}{(uniform)} & \multicolumn{2}{c||}{rationalizability} & \multicolumn{2}{c||}{level-$k$ (uniform)} & \multicolumn{2}{c||}{rationalizability} \\ \hline \hline
Level & Player 1 & Player 2 & Player 1 & Player 2 & Player 1 & Player 2 & Player 1 & Player 2 & Player 1 & Player 2 \\ \hline \hline
$1$ & $\{(O_1, *)\}$ & $\{(C_2, o_2)\}$ & $\{(O_1, *)\}$ & $\{(C_2, o_2)\}$ & $\{(O_1, *), (C_1, c_1)\}$ & $\{(O_2, *), (C_2, o_2)\}$ & $\{(O_1, c_1)\}$ & $\{(C_2, o_2)\}$ & $\{(C_1, c_2), (O_1, o_1)\}$ & $\{(*, o_2)\}$ \\ 
$2$ & $\{(O_1, *)\}$ & $\{(C_2, o_2)\}$ & $\{(O_1, *)\}$ & $S_2$ & $\{(O_1, *), (C_1, c_1)\}$ & $\{(C_2, o_2)\}$ & $\{(O_1, o_1)\}$ & $\{(C_2, o_2)\}$ & $\{(O_1, o_1)\}$ & $\{(*, o_2)\}$ \\ 
$3$ & $\{(O_1, *)\}$ & $\{(C_2, o_2)\}$ & $\{(O_1, *), (C_1, c_1)\}$ & $S_2$ & $\{(O_1, *), (C_1, c_1)\}$ & $\{(C_2, o_2)\}$ & $\{(O_1, o_1)\}$ & $\{(C_2, o_2)\}$ & $\{(O_1, o_1)\}$ & $\{(O_2, o_2)\}$ \\
$4$ & $\{(O_1, *)\}$ & $\{(C_2, o_2)\}$ & $\{(O_1, *), (C_1, c_1)\}$ & $S_2$ & $\{(O_1, *), (C_1, c_1)\}$ & $\{(C_2, o_2)\}$ & $\{(O_1, o_1)\}$ & $\{(C_2, o_2)\}$ & $\{(O_1, o_1)\}$ & $\{(O_2, o_2)\}$ \\
$5$ & $\vdots$ & $\vdots$ & $\vdots$ & $\vdots$ & $\vdots$ & $\vdots$ & $\vdots$ & $\vdots$ & $\vdots$ & $\vdots$ \\ \hline \hline
\multicolumn{11}{c}{} \\
\multicolumn{11}{c}{} \\
\hline \hline HMS & \multicolumn{2}{c||}{Strong level-$k$} & \multicolumn{2}{c||}{Level-$k$ in the normal form} & \multicolumn{2}{c||}{$k$-level strong} & \multicolumn{2}{c||}{Backward} & \multicolumn{2}{c||}{$k$-level backward} \\
game & \multicolumn{2}{c||}{(uniform)} & \multicolumn{2}{c||}{(uniform)} & \multicolumn{2}{c||}{rationalizability} & \multicolumn{2}{c||}{level-$k$ (uniform)} & \multicolumn{2}{c||}{rationalizability} \\ \hline \hline
Level & Player 1 & Player 2 & Player 1 & Player 2 & Player 1 & Player 2 & Player 1 & Player 2 & Player 1 & Player 2 \\ \hline \hline
$1$ & $\{(O_1, *), (C_1, c_1)\}$ & $\{(O_2, *)\}$ & $\{(O_1, *), (C_1, c_1)\}$ & $\{(O_2, *)\}$ & $\{(O_1, *), (C_1, c_1)\}$ & $S_2$ & $\{(*, c_1)\}$ & $\{(O_2, *)\}$ & $\{(*, c_1)\}$ & $S_2$ \\
$2$ & $\{(O_1, *), (C_1, c_1)\}$ & $\{(C_2, *)\}$ & $S_1$ & $\{(C_2, *)\}$ & $\{(O_1, *), (C_1, c_1)\}$ & $\{(C_2, *)\}$ & $\{(*, c_1)\}$ & $\{(C_2, *)\}$ & $\{(*, c_1)\}$ & $\{(C_2, *)\}$ \\
$3$ & $\{(O_1, *), (C_1, c_1)\}$ & $\{(C_2, *)\}$ & $\{(O_1, *), (C_1, c_1)\}$ & $S_2$ & $\{(O_1, *), (C_1, c_1)\}$ & $\{(C_2, *)\}$ & 
$\{(*, c_1)\}$ & $\{(C_2, *)\}$ & $\{(*, c_1)\}$ & $\{(C_2, *)\}$ \\
$4$ & $\{(O_1, *), (C_1, c_1)\}$ & $\{(C_2, *)\}$ & $\{(O_1, *), (C_1, c_1)\}$ & $\{(C_2, *)\}$ & $\{(O_1, *), (C_1, c_1)\}$ & $\{(C_2, *)\}$ & $\{(*, c_1)\}$ & $\{(C_2, *)\}$ & $\{(*, c_1)\}$ & $\{(C_2, *)\}$ \\
$5$ & $\vdots$ & $\vdots$ & $\vdots$ & $\vdots$ & $\vdots$ & $\vdots$ & $\vdots$ & $\vdots$ & $\vdots$ & $\vdots$ \\ \hline \hline
\multicolumn{11}{l}{$*$ refers to any action at the player's second information set.} \\
\multicolumn{11}{c}{} \\
\multicolumn{11}{c}{} \\
\end{tabular}
\begin{tabular}{||c||c|c||c|c||c|c||c|c||c|c||} \hline \hline Centipede & \multicolumn{2}{c||}{Strong level-$k$} & \multicolumn{2}{c||}{Backward induction} & \multicolumn{2}{c||}{$k$-level strong} & \multicolumn{2}{c||}{Backward} & \multicolumn{2}{c||}{$k$-level backward} \\
game & \multicolumn{2}{c||}{(uniform)} & \multicolumn{2}{c||}{} & \multicolumn{2}{c||}{rationalizability} & \multicolumn{2}{c||}{level-$k$ (uniform)} & \multicolumn{2}{c||}{rationalizability} \\ \hline \hline
Level & Player 1 & Player 2 & Player 1 & Player 2 & Player 1 & Player 2 & Player 1 & Player 2 & Player 1 & Player 2 \\ \hline \hline
$1$ & $\{(C_1, c_1)\}$ & $\{(C_2, o_2)\}$ & $S_1$ & $\{(O_2, o_2), (C_2, o_2)\}$ & $S_1$ & $\{(O_2, *), (C_2, o_2)\}$ & $\{(C_1, c_1)\}$ & $\{(C_2, o_2)\}$ & $S_1$ & $\{(*, o_2)\}$ \\
$2$ & $\{(C_1, o_1)\}$ & $\{(C_2, o_2)\}$ & $\{(O_1, o_1), (C_1, o_1)\}$ & $\{(O_2, o_2), (C_2, o_2)\}$ & $\{(O_1, *), (C_1, o_1)\}$ & $\{(O_2, *), (C_2, o_2)\}$ & $\{(C_1, o_1)\}$ & $\{(C_2, o_2)\}$ & $\{(*, o_1)\}$ & $\{(*, o_2)\}$ \\
$3$ & $\{(C_1, o_1)\}$ & $\{(O_2, *)\}$ & $\{(O_1, o_1), (C_1, o_1)\}$ & $\{(O_2, o_2)\}$ & $\{(O_1, *), (C_1, o_1)\}$ & $\{(O_2, *)\}$ & $\{(C_1, o_1)\}$ & $\{(O_2, o_2)\}$ & $\{(*, o_1)\}$ & $\{(O_2, o_2)\}$ \\
$4$ & $\{(O_1, *)\}$ & $\{(O_2, *)\}$ & $\{(O_1, o_1)\}$ & $\{(O_2, o_2)\}$ & $\{(O_1, *)\}$ & $\vdots$ & $\{(O_1, o_1)\}$ & $\{(O_2, o2)\}$ & $\{(O_1, o_1)\}$ & $\{(O_2, o_2)\}$ \\
$5$ & $\vdots$ & $\vdots$ & $\vdots$ & $\vdots$ & $\vdots$ & $\vdots$ & $\vdots$ & $\vdots$ & $\vdots$ & $\vdots$ \\ \hline \hline
\multicolumn{11}{l}{$*$ refers to any action at the player's second information set.} \\
\multicolumn{11}{c}{} \\
\multicolumn{11}{c}{} \\
\end{tabular}
\end{sidewaystable}

The observation does not extend to level-$2$ thinking strategies as the following example demonstrates:

\begin{ex}\label{Reny_example}\textbf{\emph{(Strong level-$k$ strategies refine level-$k$ strategies in the normal form for $k \geq 2$)}} Consider a version of a game in Figure~\ref{Reny} due to Reny (1992) (which is a variant of a centipede game). To the left, we print the game in extensive form; to the right the associated normal form.\footnote{Ignore for the time being the red lines and blue boxes in the normal form of Figure~\ref{Reny}. They pertain to strong rationalizability and will be discussed later.}
\begin{figure}[h!]\caption{Reny (1992) Game\label{Reny}}
	\begin{center}
	\includegraphics[scale=0.4]{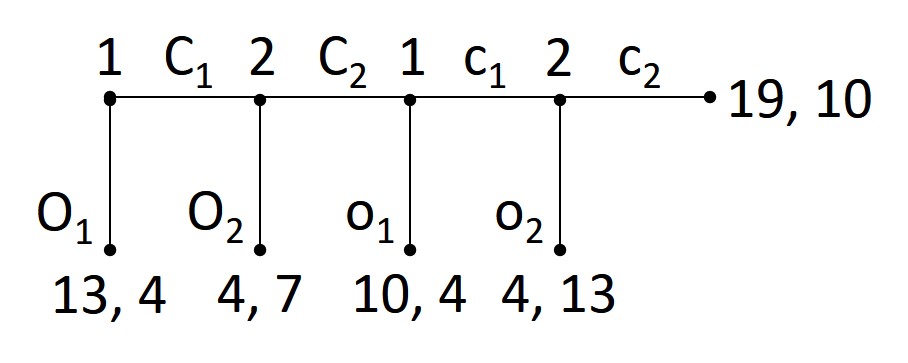} 	\includegraphics[scale=0.4]{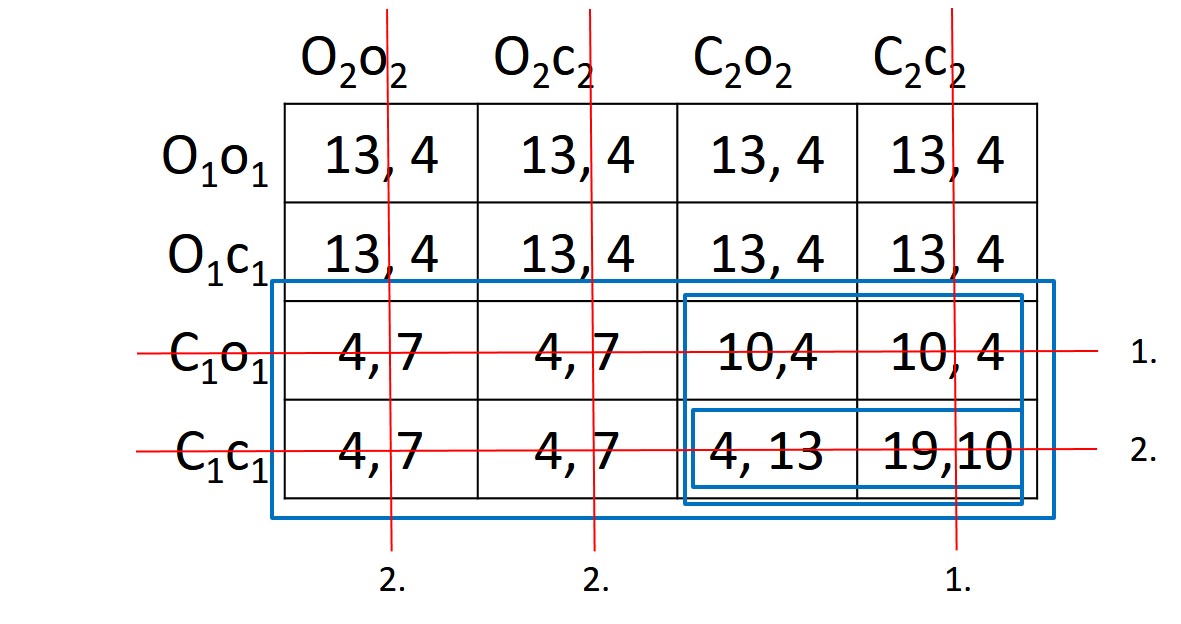}
    \end{center}
\end{figure} The strong level-$k$ thinking strategies and the level-$k$ thinking strategies in the normal form are printed in Table~\ref{solutions_table} for any $k \geq 1$. (We let ``$*$'' stand for any action.) We observe that at any level $k \geq 2$, the strong level-$k$ strategies of player 2 are a subset of level-$k$ thinking strategies in the normal form. For $k \geq 3$, this holds not only for strategies of player 2 but even outcomes. The reason is that once player 2 gets to move (i.e., reaches her first information set), she is certain at level 2 that player 1 does not use strong level-$1$ thinking since any strong level-$1$ thinking strategy of player 1 prescribes $O_1$ at the root of the tree. She must now think that player 1 behaves uniformly over her strategies and with such a belief her strategy $(C_2, o_2)$ is uniquely rational. In contrast, level-$2$ thinking in the normal form of player 2 presumes level-$1$ thinking strategies of player 1, i.e., any strategy in $\{(O_1, o_1), (O_1, c_1)\}$. Thus, player 2 is indifferent among all her strategies and hence any strategy in $S_2$ is consistent with level-$2$ thinking in the normal form.
\end{ex}

Above example shows that at level $2$, only strategies differ between strong level-$k$ thinking and level-$k$ thinking in the normal form but not outcomes. One could argue that this fact is behaviorally irrelevant at level $2$ as only outcomes ``should'' matter. Note though that strategies could be elicited via the strategy method (Selten, 1967) so that not just outcomes of the game in extensive form are behaviorally relevant. Nevertheless, even if one takes the view that only outcomes matter, it should be noted that in this example outcomes differ at level $k \geq 3$ as well. Moreover, the next example shows that already at level $2$, outcomes of strong level-$k$ thinking can differ from level-$k$ thinking in the normal form.

\begin{ex}\label{HMS_example}\textbf{\emph{(Strong level-$2$ outcomes refine level-$2$ thinking outcomes in the normal form)}} Consider the game in Figure~\ref{HMS} from Heifetz, Meier, and Schipper (2021) (i.e., HMS game for short):
\begin{figure}[h!]\caption{HMS Game\label{HMS}}
	\begin{center}
	\includegraphics[scale=0.4]{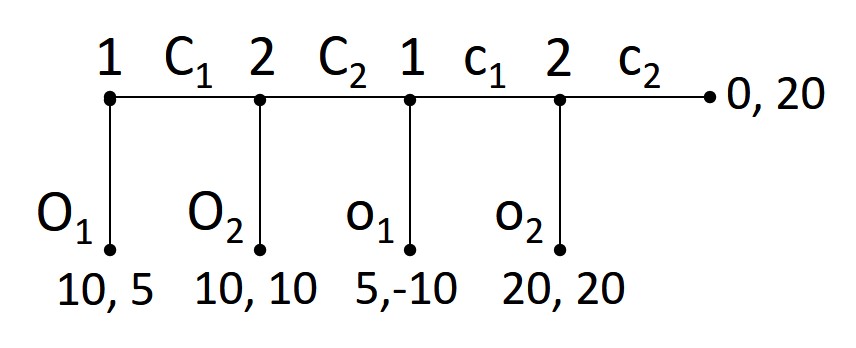} 	\includegraphics[scale=0.4]{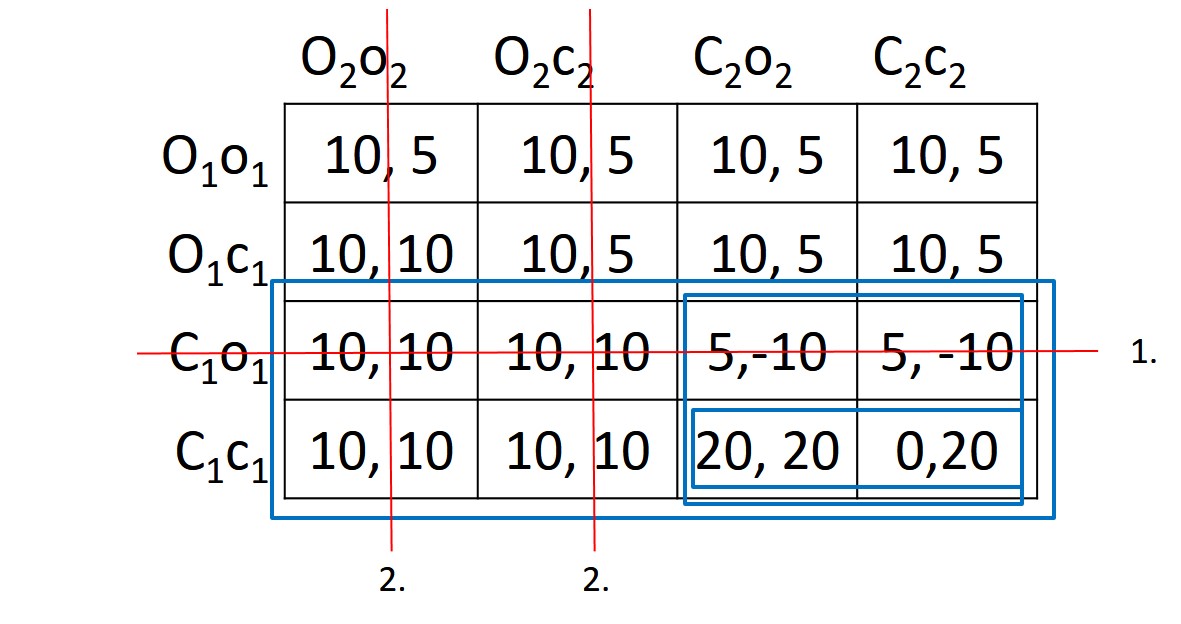}
    \end{center}
\end{figure} The strong level-$k$ thinking strategies and the level-$k$ thinking strategies in the normal form (both with uniform level-$1$ beliefs) are printed in Table~\ref{solutions_table} for any $k \geq 1$. We observe that at level $2$, the strong level-$2$ thinking strategies are $\{(O_1, *), (C_1, c_1)\}$ and $\{(C_2, *)\}$ for players 1 and 2, respectively, while for level-$2$ thinking in the normal form they are $S_1$ and $\{(C_2, *)\}$ for players 1 and 2, respectively. (Again, we let ``$*$'' stand for any action.) In particular, level-$2$ thinking in the normal form allows the outcome with payoffs $(5, -10)$ to emerge while this is ruled out with strong level-$2$ thinking. The reason is that under level-$2$ thinking in the normal form, player $1$ is indifferent among all strategies in the normal form since he believes that player $2$ plays $O_2$. In contrast, in the extensive form, when player $1$ reaches his second information set, he is certain that player $2$ does not follow strong level-$1$ thinking but must be ``level-$0$''. Consequently, at his second information set, he has uniform beliefs about the actions of player $2$ at the last information set, with which only $c_1$ is rational.
\end{ex}

Both examples beg the question of whether or not in general for any level $k \geq 1$, strong level-$k$ thinking is a (weak) refinement (both in terms of strategies and outcomes) of level-$k$ thinking in the normal form. The examples suggest this to be the case. More generally, we can show that for any full-support initial belief system $\bar{\beta}^1$ and level $k$, the set of outcomes reached by strong level-$k$ thinking refines the set of outcomes reached by level-$k$ thinking in the normal form.

To state this assertion more formally, we require the following definition. For any strategy profile $s \in S$, let $z(s)$ denote the terminal history reached by $s$. For any nonempty subset of strategy profiles $S' \subseteq S$, let $Z(S') = \{z \in Z : z = z(s), s \in S'\}$. Note that for any nonempty subsets $S', S''$ of $S$, $S' \subseteq S''$ implies $Z(S') \subseteq Z(S'')$.

\begin{prop}\textbf{\emph{(Strong level-$k$ thinking refines outcomes of level-$k$ thinking in the normal form)}}\label{refinement} Consider any finite game in extensive form with perfect recall. Let $\bar{\beta}^1$ be a profile of full-support belief systems and $\beta^1$ a profile of full-support beliefs in the associated normal form consistent with $\bar{\beta}^1$. Then $Z(\bar{L}^k(\bar{\beta}^1)) \subseteq Z(L^k(\beta^1))$ for all $k \geq 1$.
\end{prop}

The proof proceeds by induction. The result for the base-case, level-$1$, is implied by Proposition~\ref{EFL1_NFL1}. At any higher level-$k$, we focus on players whose information set is reached along the path to the outcome noting that for other players we can simply select a strong level-$(k-1)$ thinking strategy without affecting the outcome. For any player along the path, we select the first information set, which is well-defined since the game as perfect recall. The strong level-$k$ rational strategy is also rational at this information set with a belief that  - as we show - we can confine to level-$(k-1)$ strategies of opponents in the normal form. The result now follows.

The proposition implies in particular that, if initial beliefs are uniform, as often assumed in applications, then for any level $k$, strong level-$k$ thinking refines the set of outcomes that can be reached by level-$k$ thinking in the normal form.

We note that the games used in Subsections~\ref{Cooper_section} and~\ref{BN_section} show that for some games the outcome-refinement of strong level-$k$ thinking is strict at some levels.

\subsection{Strong Level-$k$ Thinking versus Strong Rationalizability}

In Section~\ref{NF_levelk_rat} we observed that any level-$k$ thinking strategy in the normal form is also level-$k$ rationalizable. In this section, we show that this is not necessarily the case anymore when we consider strong level-$k$ thinking and level-$k$ strong rationalizable strategies. It is not due to a defect in the definitions. Rather, it is due to different updates conditional on information sets that may occur with and without restrictions on first-order belief systems of strong level-$k$ thinking.

The notion of strong rationalizability is also due to Pearce (1984), who called it rationalizability in the extensive form. In contrast to its well-known counterpart for games in normal form, there is no treatment of strong rationalizability in standard textbooks on game theory. Consequently, it is much less known although undeservingly so. We follow Battigalli (1997) in allowing for correlated beliefs over opponents' strategies and define it as a reduction procedure on beliefs that subsequently implies a reduction procedure on strategies. The following definition is due to Battigalli (1997), who originally called it extensive-form rationalizability and proved it to be equivalent to Pearce's definition allowing for correlated beliefs. In private communication, Pierpaolo Battigalli ``strongly'' encouraged us to now change the terminology to ``strong rationalizability''. 

\begin{defin}[Strong Rationalizability]\label{EFR} Define inductively for every player $i \in N$,
\begin{eqnarray*}	
\bar{B}_i^1 & & \mbox{is the set of player $i$'s belief systems.}\\
\bar{R}_i^1 & = & \left\{ s_i \in S_i : \begin{array}{l} \mbox{There exists } \bar{\beta}_i \in \bar{B}_i^1 \mbox{ with which for every } \\ \mbox{information set } I_i \in \mathcal{I}_i, s_i \mbox{ is rational at } I_i. \end{array} \right\} \\
& \vdots & \\
\bar{B}_i^k & = & \left\{ \bar{\beta}_i \in \bar{B}_i^{k - 1} : \begin{array}{l} \mbox{For every information set } I_i, \mbox{ if there exists some profile of } \\
\mbox{other players' strategies } s_{-i} \in \bar{R}_{-i}^{k - 1}
\mbox{ such that } s_{-i} \mbox{ reaches } I_i, \\ \mbox{then }
\bar{\beta}_i(I_i) \mbox{ assigns probability } 1 \mbox{ to } \bar{R}_{-i}^{k - 1}. \end{array}\right\}\\
\bar{R}_i^k & = & \left\{ s_i \in S_i : \begin{array}{l} \mbox{There exists } \bar{\beta}_i \in \bar{B}_i^k \mbox{ with which for every } \\\mbox{information set } I_i \in \mathcal{I}_i,
s_i \mbox{ is rational at } I_i. \end{array} \right\}
\end{eqnarray*} The set of strong rationalizable strategies of player $i$ is $$\bar{R}_i^{\infty} = \bigcap_{k = 1}^{\infty} \bar{R}_i^k.$$
\end{defin}

Battigalli and Siniscalchi (2002) characterize strong rationalizability by common strong belief in rationality. Moreover, for every finite level-$k$, the $k$-level strong rationalizable strategies are characterized by $k$-level mutual strong belief in rationality. Similar to the equivalence between rationalizability and iterated elimination of strictly dominated actions, strong rationalizability is level-by-level strategy-equivalent to iterated elimination of conditionally strictly dominated strategies (Shimoji and Watson, 1998). For each information set of a player, consider the sub-space of strategy profiles that reach this information set. This is a normal-form information set in the sense of Mailath, Samuelson, and Swinkels (1993). A strategy is conditionally strictly dominated if there exists a possibly mixed strategy that conditional on such a normal-form information set yields a strictly higher expected utility no matter what strategy profiles in the normal-form information set are played by opponents.

In the figures, we indicate in the associated normal form the normal-form information sets associated to information sets in the extensive form by blue boxes. We also indicate the order of elimination of conditionally strictly dominated strategies (and hence the order of elimination of non-rationalizable strategies) by red lines with numbers attached that represent the level at which the strategy is eliminated. These examples illustrate already some findings. In the Reny game (Figure~\ref{Reny}), strong level-$k$ thinking strategies refine level-$k$ strong rationalizable strategies for $k \geq 1$. In the HMS game (Figure~\ref{HMS}), they refine strong rationalizable strategies of player 2 at level $1$ but otherwise yield the same strategies at higher levels. For both games, the strong level-$k$ thinking strategies (with uniform level-$1$ belief systems) and $k$-level strong rationalizable strategies are printed level-by-level in Table~\ref{solutions_table}. (Again, we let ``$*$'' stand for any action.)
\begin{figure}[h!]\caption{Battle-of-the-sexes with an outside option I (BoS I)\label{BoS1}}
	\begin{center}
    \includegraphics[scale=0.4]{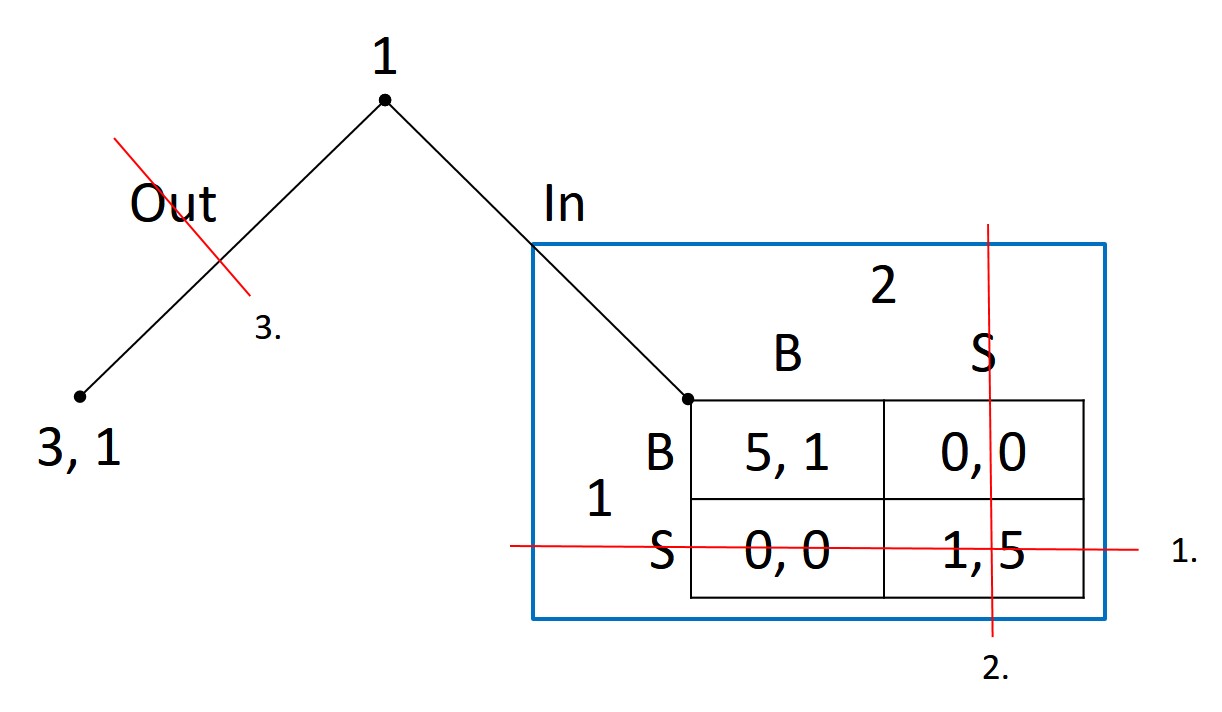}
    \end{center}
\end{figure} 
\begin{table}[h!]\caption{Solutions to Outside Option Games\label{BoS_table}}
\tiny
\centering
\begin{tabular}{||c||c|c||c|c||c|c||c|c||} \hline \hline BoS I & \multicolumn{2}{c||}{Strong level-$k$} & \multicolumn{2}{c||}{$k$-level strong} & \multicolumn{2}{c||}{Backward} & \multicolumn{2}{c||}{$k$-level backward}\\
game & \multicolumn{2}{c||}{(uniform)} & \multicolumn{2}{c||}{rationalizability} & \multicolumn{2}{c||}{level-$k$ (uniform)} & \multicolumn{2}{c||}{rationalizability} \\ \hline \hline
Level & Player 1 & Player 2 & Player 1 & Player 2 & Player 1 & Player 2 & Player 1 & Player 2 \\ \hline \hline
$1$ & $\{(Out, *)\}$ & $\{S\}$ & $\{(Out, *), (In, B)\}$ & $\{B, S\}$ & $\{(Out, B)\}$ & $\{S\}$ & $\{(Out, S), (In, B)\}$ & $\{B, S\}$ \\
$2$ & $\{(Out, *)\}$ & $\{S\}$ & $\{(Out, *), (In, B)\}$ & $\{B\}$ & $\{(Out, B)\}$ & $\{B\}$ & $\{(Out, S), (In, B)\}$ & $\{B, S\}$ \\
$3$ & $\vdots$ & $\vdots$ & $\{(In, B)\}$ & $\vdots$ & $\{(In, B)\}$ & $\{B\}$ & $\{(Out, S), (In, B)\}$ & $\{B, S\}$ \\ \hline \hline
\multicolumn{9}{c}{} \\
\multicolumn{9}{c}{} \\
\hline \hline BoS II & \multicolumn{2}{c||}{Strong level-$k$} & \multicolumn{2}{c||}{$k$-level strong} & \multicolumn{2}{c||}{Backward} & \multicolumn{2}{c||}{$k$-level backward}\\
game & \multicolumn{2}{c||}{(uniform)} & \multicolumn{2}{c||}{rationalizability} & \multicolumn{2}{c||}{level-$k$ (uniform)} & \multicolumn{2}{c||}{rationalizability} \\ \hline \hline
Level & Player 1 & Player 2 & Player 1 & Player 2 & Player 1 & Player 2 & Player 1 & Player 2 \\ \hline \hline
$1$ & $\{(In, B)\}$ & $\{S\}$ & $\{(Out, *), (In, B)\}$ & $\{B, S\}$ & $\{(In, B)\}$ & $\{S\}$ & $\{(Out, S), (In, B)\}$ & $\{B, S\}$ \\
$2$ & $\{(Out, *)\}$ & $\{B\}$ & $\{(Out, *), (In, B)\}$ & $\{B\}$ & $\{(Out, S)\}$ & $\{B\}$ & $\{(Out, S), (In, B)\}$ & $\{B, S\}$ \\
$3$ & $\{(In, B)\}$ & $\{B\}$ & $\{(In, B)\}$ & $\vdots$ & $\{(In, B)\}$ & $\{S\}$ & $\{(Out, S), (In, B)\}$ & $\{B, S\}$ \\
$4$ & $\vdots$ & $\vdots$ & $\vdots$ & $\vdots$ & Cycle & Cycle & $\vdots$ & $\vdots$ \\ \hline \hline
\multicolumn{9}{c}{} \\
\multicolumn{9}{c}{} \\
\end{tabular}
\normalsize
\end{table}

One of the most prominent examples to demonstrate the forward induction power of strong rationalizability is the battle-of-the-sexes game with an outside option. We use this example to discuss the relationship between strong level-$k$ thinking and level-$k$ strong rationalizability. In particular, we show that in contrast to the analogous normal-form solution concepts, strong level-$k$ thinking does not refine level-$k$ strong rationalizability.

\begin{ex}\label{BoSI_example} \textbf{\emph{(Battle-of-the-sexes with an outside option I)}} Consider the game in Figure~\ref{BoS1}. Player 1 moves first, deciding between $Out$ and $In$. When he chooses $In$, the battle-of-the-sexes game is played. Both players can choose between $B$ and $S$. Player 1 strictly prefers $((In, B), B)$ over $((Out, *), *)$ over $((In, S), S)$ and any other outcome. While for levels $k \leq 2$, strong level-$k$ thinking with uniform initial beliefs refines level-$k$ strong rationalizable outcomes, at level $3$ and higher the strong level-$k$ thinking outcomes are distinct from the level-$k$ strong rationalizable outcomes, respectively (see Table~\ref{BoS_table}, upper part). For player 1 at level 1, $(Out, *)$ is rational given a uniform belief over player 2's strategies.\footnote{Also Balkenborg and Nagel (2016, p. 398) present such an argument for moving Out at the beginning of the game.} For strong rationalizability, also $(In, B)$ is rational at level 1. Consequently, if player 2's information set is reached at level 2, she knows that player 1 continues with action $B$ to which $B$ is the unique best response by player 2. In contrast, strong level-1 thinking strategies (with uniform initial belief systems) do not reach player 2's information set. Hence, when she is called to play, she must believe that player 1 is level-0, i.e., playing uniformly. With such a belief, $S$ is uniquely rational for player 2. At level 2, player 1 knows that player 2 would play $S$ upon moving $In$. Given this belief, $(Out, *)$ continues to be rational at the level 2. That is, the prediction of strong level-$3$ is $(Out, *)$. In contrast, for strong rationalizability, player 1 understands at level 3 that once she moves $In$, player 2 will select $B$. Thus, playing $B$ upon moving $In$ is the unique best response of player 1. To summarize, strong level-$3$ thinking strategies with uniform initial belief yield $((Out, *), S)$ while $3$-level strong rationalizability strategies yield a different outcome, $((In, B), B)$. This example demonstrates that generally strong level-$k$ thinking with uniform belief systems is neither a refinement nor is refined by $k$-level strong rationalizability.
\end{ex}

The outcome of strong level-$k$ thinking in the prior example is driven by the fact that with player 1's uniform belief over player 2's actions, $(Out, *)$ is rational, i.e., yielding a utility of 3, which is larger than 2.5, the expected utility from playing $(In, B)$ against an uniformly mixing player 2. This observation illustrates that outcomes of strong level-$k$ thinking can be sensitive to misspecifications of utilities, which poses interesting issues in experiments. Obviously, the non-robustness is problematic because we can only experimentally control payoffs but not necessarily utilities. On the other hand, small changes in the utility matter behaviorally in a way that is not picked up by $k$-level strong rationalizability but is picked up by strong level-$k$ thinking because of restrictions on first-level belief systems. This yields testable predictions of strong level-$k$ thinking if payoffs are taken at face value as utilities in experiments. The prediction of strong level-$k$ thinking can change radically when slightly changing player $1$'s utility of the outside option in a way that does not affect predictions of strong rationalizability. This is demonstrated in the next example.
\begin{figure}[h!]\caption{Battle-of-the-sexes with an outside option II\label{BoS2}}
	\begin{center}
    \includegraphics[scale=0.4]{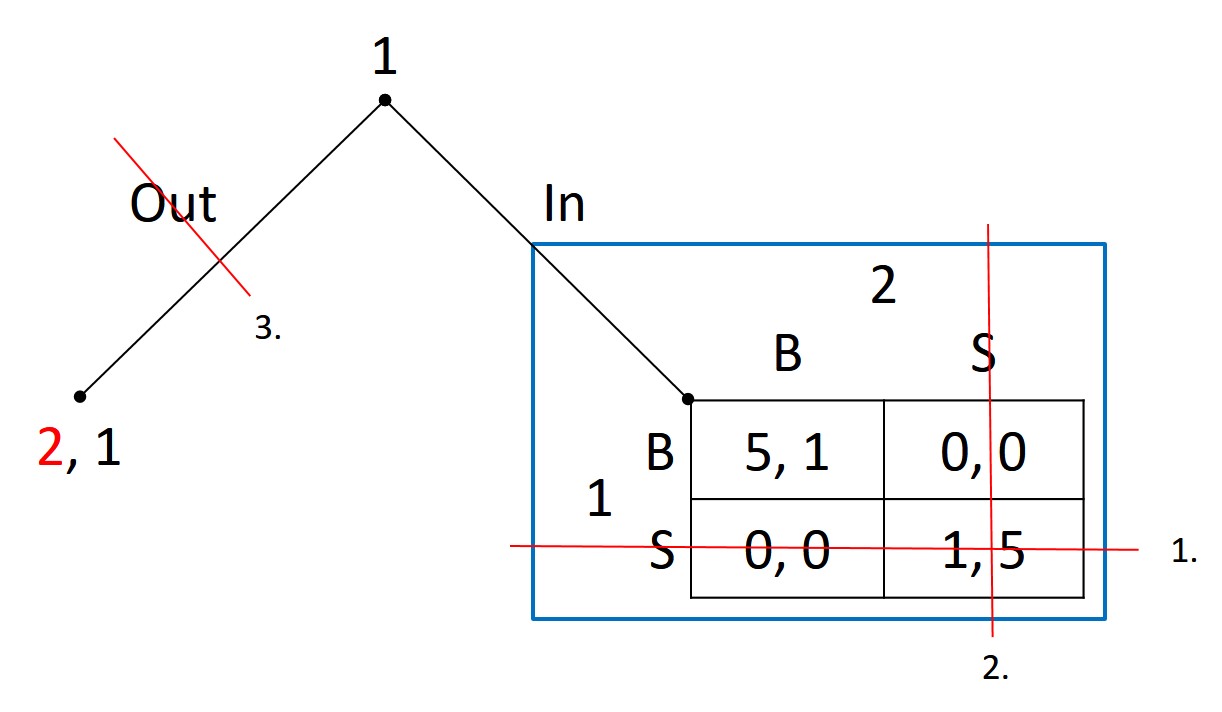}
    \end{center}
\end{figure}

\begin{ex}\label{BoSII_example} \textbf{\emph{(Battle-of-the-sexes with an outside option II)}} Consider the game in Figure~\ref{BoS2}. This game is identical to the game in Figure~\ref{BoS1} expect that the outside option yields now a utility of 2 instead of 3 to player 1. The predictions of $k$-level strong rationalizability remain unchanged and are level-by-level identical to the ones for the game in Figure~\ref{BoS1}. Yet, strong level-$k$ thinking strategies (with uniform initial belief systems) of player 1 differ considerably from the ones in Figure~\ref{BoS1} (see Table~\ref{BoS_table}, lower part). For player 1 at level 1, $(In, B)$ is uniquely rational at both of player $1$'s information sets to uniform beliefs over player 2's strategies. Player 2 plays optimally $S$ against an uniformly mixing player 1. At the second level, player 1 anticipates this and $(Out, *)$ is uniquely rational with such a belief. For player 2, $B$ is the only rational action at level 2 because she knows by now that once her information set is reached, player 1 continues to play $B$. Anticipating this at level 3, player 1 selects $(In, B)$. Player 2 continues with $B$ at level 3 because strong level-$2$ thinking with uniform initial belief systems of player 1 prescribes $(Out, *)$. That is, her information set should not be reached with such a strategy. Upon reaching the information set nevertheless, she must believe that player 1 is strong level-$1$, i.e., playing $(In, B)$, and consequently $B$ is the unique best response. Thus, strong level-$k$ thinking yields the strong rationalizable outcome and strategies for $k \geq 3$. The point of the example is to demonstrate that changing the utility of the outside option to player 1 from 3 to 2 alters dramatically strong level-$k$ thinking strategies. More generally, given uniform initial beliefs of player 1, the prediction flips when player 1's utility of the outside option crosses 2.5. In contrast, this change does not affect the prediction of strong rationalizability.
\end{ex}

To sum up, strong level-$k$ thinking with uniform initial belief systems and strong rationalizability are unrelated solution concepts in terms of outcomes, which is in contrast to their normal-form counterparts (see Section~\ref{NF_levelk_rat}). This is due to how restrictions placed on first-level belief systems (i.e., level-$0$ behavior) interact with conditional beliefs. It also makes strong level-$k$ thinking very sensitive to misspecifications of utilities.

\subsection{Strong Level-$k$ Thinking versus Strong $\Delta$-Rationalizability\label{Delta_section}}

In the prior subsection, we argued that strong level-$k$ thinking differs from strong rationalizability due to restrictions placed on first-level belief systems (aka level-$0$ behavior). Yet, in the literature there are already versions of strong rationalizability that incorporate restrictions on first-level beliefs under the name of strong $\Delta$-rationalizability (see Battigalli, 2003, Battigalli and Siniscalchi, 2003, Battigalli and Prestipino, 2013) and extensive-form best response sets (Battigalli and Friedenberg, 2012). Here we state the definition of strong $\Delta$-rationalizability in a form that facilitates the comparison with strong rationalizability and strong level-$k$ thinking.

\begin{defin}[Strong $\Delta$-Rationalizability]\label{DeltaR} For each player $i \in N$, fix a nonempty (measurable) set of restrictions on belief systems $\Delta_i \subseteq \bar{B}_i$. Let $\Delta := (\Delta_i)_{i \in N}$. Define inductively for every player $i \in N$,
\begin{eqnarray*}	
\bar{B}_i^1(\Delta) & & \mbox{is the set of player $i$'s restricted belief systems $\Delta_i$.}\\
\bar{R}_i^1(\Delta) & = & \left\{ s_i \in S_i : \begin{array}{l} \mbox{There exists } \bar{\beta}_i \in \bar{B}_i^1(\Delta) \mbox{ with which for every} \\ \mbox{information set } I_i \in \mathcal{I}_i, s_i \mbox{ is rational at } I_i. \end{array} \right\} \\
& \vdots & \\
\bar{B}_i^k(\Delta) & = & \left\{ \bar{\beta}_i \in \Delta_i : \begin{array}{l} \mbox{For every information set } I_i, \mbox{ if there exists some profile of } \\
\mbox{other players' strategies } s_{-i} \in \bar{R}_{-i}^{k - 1}(\Delta)
\mbox{ such that } s_{-i} \mbox{ reaches } I_i, \\ \mbox{then }
\bar{\beta}_i(I_i) \mbox{ assigns probability } 1 \mbox{ to } \bar{R}_{-i}^{k - 1}(\Delta). \end{array}\right\}\\
\bar{R}_i^k(\Delta) & = & \left\{ s_i \in \bar{R}_{i}^{k - 1}(\Delta) : \begin{array}{l} \mbox{There exists } \bar{\beta}_i \in \bar{B}_i^k(\Delta) \mbox{ with which for every} \\\mbox{information set } I_i \in \mathcal{I}_i,
s_i \mbox{ is rational at } I_i. \end{array} \right\}
\end{eqnarray*} The set of strong $\Delta$-rationalizable strategies of player $i$ is $$\bar{R}_i^{\infty}(\Delta) = \bigcap_{k = 1}^{\infty} \bar{R}_i^k(\Delta).$$
\end{defin}

First note that compared to strong level-$k$ thinking, strong $\Delta$-rationalizability is a reduction procedure on strategies. Yet, it is also easy to see that the set of $\Delta$-rationalizability strategies may be empty. This is naturally the case when the restrictions $\Delta_i$ clash with the requirement that beliefs must assign probability 1 to $\bar{R}_{-i}^{k - 1}(\Delta)$ when latter set is not ruled out. That is, the crux is in the definition of $\bar{B}_i^k(\Delta)$. We require $\bar{\beta}_i \in \Delta_i$ which might be inconsistent with $\bar{\beta}_i(I_i)(\bar{R}_{-i}^{k - 1}(\Delta)) = 1$ and the existence of a strategy profile $s_{-i} \in \bar{R}_{-i}^{k - 1}(\Delta)$ that reaches $I_i$. There are restrictions where the set of $\Delta$-rationalizable strategies is nonempty (see Battigalli, 2003, for non-trivial applications). For instance, in the case of no restrictions, strong $\Delta$-rationalizability is equivalent to strong rationalizability and hence nonempty for every finite game. Most relevant for our comparison with strong level-$k$, the set of strong $\Delta$-rationalizable outcomes is typically empty if $\Delta_i$ is the set of full support belief systems or, even more special, the belief system of uniform beliefs (as often assumed in the level-$k$ literature). We illustrate this in the next example.
\begin{figure}[h!]\caption{Battle-of-the-sexes with an outside option III\label{BoS3}}
	\begin{center}
    \includegraphics[scale=0.4]{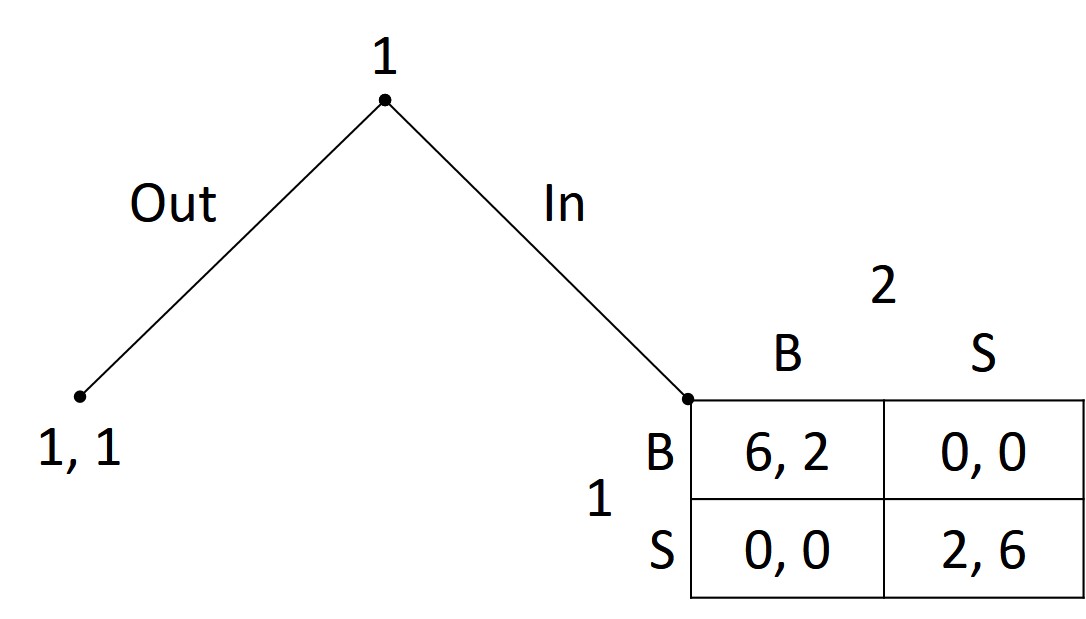}
    \end{center}
\end{figure}

\begin{ex}\label{BoSIII_example}\textbf{\emph{(Battle-of-the-sexes with an outside option III)}} Consider the game in Figure~\ref{BoS3}, a version of a game used in experiments by Cooper et al. (1993). The outside option yields now a payoff strictly lower than any pure equilibrium payoffs in the battle-of-the-sexes game. Consider $\Delta = (\Delta_1, \Delta_2)$ where $\Delta_1$ consists only of the uniform belief system, i.e., player 1's belief at the beginning of the game and after moving $In$ assigns assigns probability $\frac{1}{2}$ to each action, $B$ and $S$, of player 2. Similarly, $\Delta_2$ consist of the uniform belief $\left(\frac{1}{2}, \frac{1}{2}\right)$ of player 2 over player 1's actions $B$ and $S$ after observing $In$. The strong $\Delta$-rationalizable strategies of the players 1 and 2 at level 1 are $(In, B)$ for player 1 and $S$ for player 2.  Now, at level 2, neither player can believe in the level-1 $\Delta$-rationalizable strategy of the opponent \emph{and} have uniform beliefs. Thus, level-2 $\Delta$-rationalizable strategies with $\Delta$ being the uniform belief restriction must be empty and so for $k$-level $\Delta$-rationalizable strategies, for any $k \geq 2$; see Table~\ref{BoSIII_solution}.\footnote{The discussion treats strong $\Delta$-rationalizability in a somewhat unfair way. We think the $\Delta$-restrictions where not intended to model uniform beliefs but rather some equilibrium conventions.} 
\end{ex}

Clearly, as the example illustrates, in the case of uniform belief restrictions, we need to give up this restriction at level-$2$ and higher. This motivates a modification of strong $\Delta$-rationalizablity in which we replace $\bar{\beta}_i \in \Delta_i$ with $\bar{\beta}_i \in \bar{B}_i$ in the specification of $\bar{B}_i^k(\Delta)$ of Definition~\ref{DeltaR}. Returning to Example~\ref{BoSIII_example}, we notice now that the ``modified'' $\Delta$-rationalizable strategies are nonempty. In particular, they coincide with the first level $\Delta$-rationalizable strategies, $(In, B)$ and $S$ for players 1 and 2, respectively. Yet, this is a very strange ``solution''. If player 1 realizes that player 2 plays $S$, why wouldn't he best respond with $(In, S)$? Similarly, if player 2 realizes that player 1 plays $(In, B)$, why would not she best respond with $B$. The reason is that we require nestedness of strategies in the specification of $R_i^k(\Delta)$ of Definition~\ref{DeltaR}, i.e., $R_i^k(\Delta) \subseteq R_i^{k-1}(\Delta)$, for $k > 1$. Giving up in the modified definition of strong $\Delta$-rationalizability this nestedness yields our definition of strong level-$k$ thinking. We derive the strong level-$k$ thinking strategies with uniform belief systems of Example~\ref{BoSIII_example} in Table~\ref{BoSIII_solution}. It illustrates that the non-nestedness of strong level-$k$ thinking may create cycles, similar to what we have already observed for normal-form level-$k$ thinking in Example~\ref{cycle}. To sum up, we can interpret strong level-$k$ thinking as a version of strong $k$-level $\Delta$-rationalizability, where first-level belief restrictions are given up at higher levels together with nestedness of beliefs across levels. We expect that this would cause problems for epistemic characterizations of the solution concept in some games. Such lack of epistemic justification may just reflect the fact that in some games it is difficult to resolve what to play with level-by-level thinking etc. as already suggested in the quote by Morgenstern (1928) in the Introduction. 

\begin{table}[h!]\caption{Solutions to the Battle-of-the-sexes game with an outside option III\label{BoSIII_solution}}
\footnotesize
\begin{center}
\begin{tabular}{||c||c|c||c|c||c|c||} \hline \hline  & \multicolumn{2}{c||}{Strong level-$k$ \&} & \multicolumn{2}{c||}{$k$-level strong \&} & \multicolumn{2}{c||}{$k$-level $\Delta$ rationalizability} \\
  & \multicolumn{2}{c||}{Backward level-$k$} & \multicolumn{2}{c||}{$k$-level backward} & \multicolumn{2}{c||}{with the $\Delta$-restriction being} \\ 
 & \multicolumn{2}{c||}{(uniform)} & \multicolumn{2}{c||}{rationalizability} & \multicolumn{2}{c||}{the uniform belief system} \\ \hline \hline
Level & Player 1 & Player 2 & Player 1 & Player 2 & Player 1 & Player 2 \\ \hline \hline
$1$ & $(In, B)$ & $S$ & $(In, *)$ & $\{B, S\}$ & $(In, B)$ & $S$\\
$2$ & $(In, S)$ & $B$ & $(In, *)$ & $\{B, S\}$ & $\emptyset$ & $\emptyset$ \\
$3$ & $(In, B)$ & $S$ & $(In, *)$ & $\{B, S\}$ & $\emptyset$ & $\emptyset$ \\
$\vdots$ & Cycle & Cycle & $\vdots$ & $\vdots$ & $\vdots$ & $\vdots$ \\ \hline \hline
\end{tabular}
\end{center}
\normalsize
\end{table}

\subsection{Strong Level-$k$ Thinking versus Iterated Admissibility}

In the battle-of-the-sexes games with an outside option of the previous section, $k$-level strong rationalizability yields strategies that are equivalent to $k$-iterative elimination of weakly dominated strategies. Since in the battle-of-the-sexes game with an outside option I, strong level-$k$ thinking with uniform initial beliefs is distinct from $k$-level strong rationalizability, it demonstrated that strong level-$k$ thinking is also distinct from $k$-iterative admissibility even if we assume uniform initial belief. Yet, especially with the assumption of initial full-support belief, it is intuitive that sometimes strong level-$k$ thinking retains some features of iterated admissibility because iterated admissibility can be thought of as rationalizability with full-support beliefs. Both, strong level-$k$ with initial full support belief systems and iterated admissibility feature some form of caution. This is indeed the case in the following example.
\begin{figure}[h!]\caption{HMS2 Game\label{HMS2}}
	\begin{center}
    \includegraphics[scale=0.3]{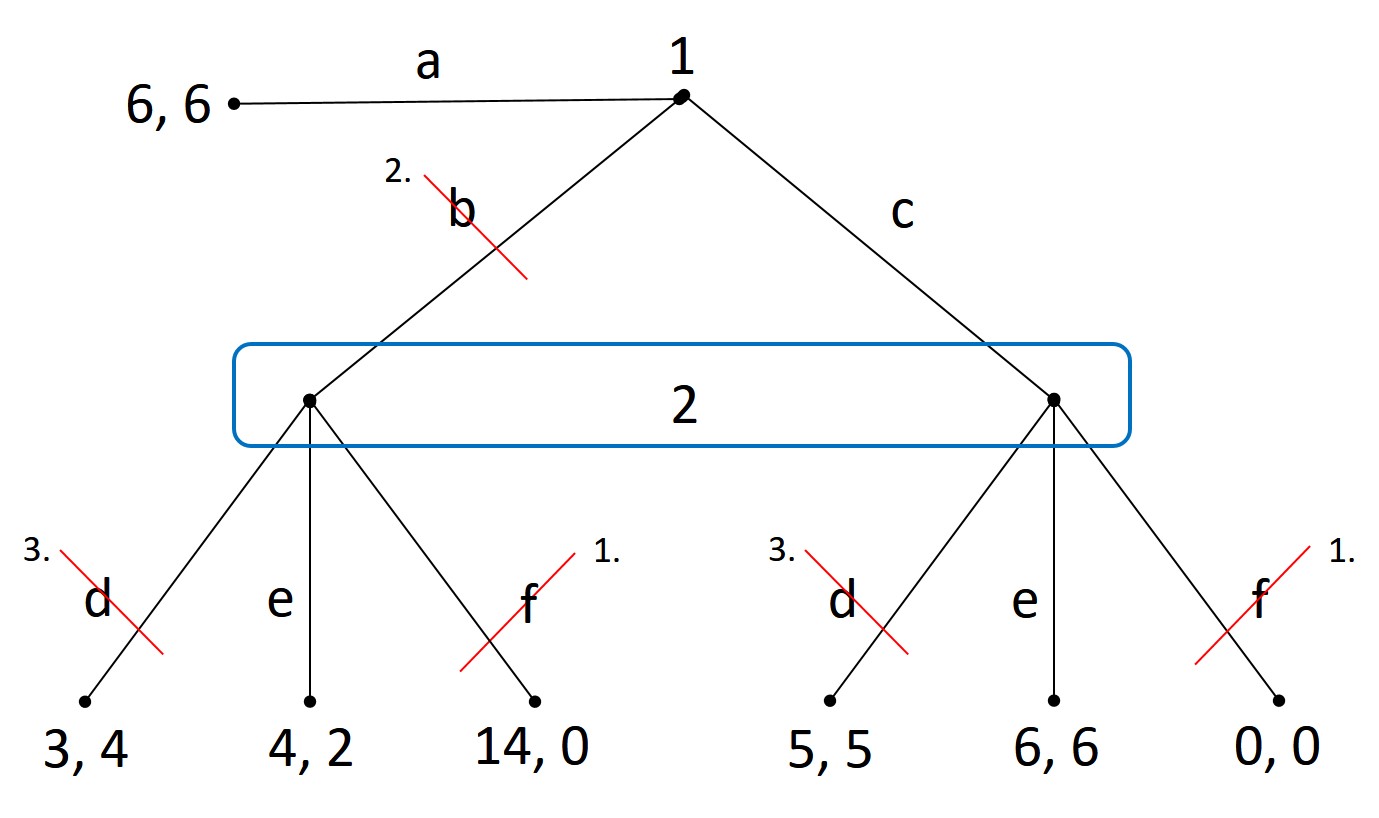} \includegraphics[scale=0.3]{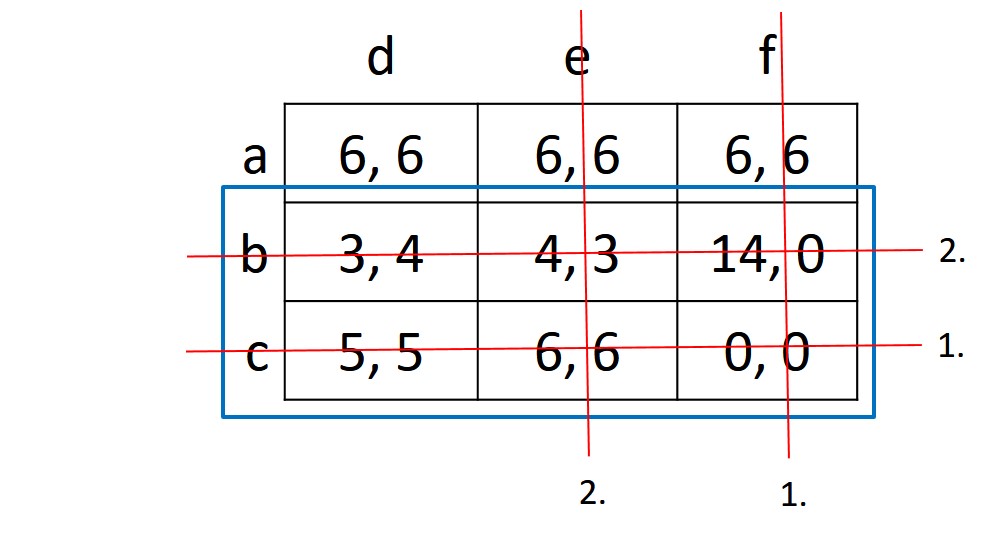}
    \end{center}
\end{figure}

\begin{ex}\label{HMS_example} Consider the game of Figure~\ref{HMS2} that is a variant of a game discussed in Heifetz, Meier, and Schipper (2021). Let's call it the HMS2 game. In this game, strong rationalizable strategies are disjoint from strategies remaining from iterated admissibility although in terms of outcomes, iterated admissibility strictly refines the set of strong rationalizable outcomes. (See Table~\ref{solutions_table_HMS23}, upper part, or the eliminations in Figure~\ref{HMS2} (for strong rationalizability) and the associated normal form (for iterated admissibility)). Strong level-$k$ thinking (with uniform initial belief systems) coincides with $k$-iterated admissibility and is disjoint from $k$-level strong rationalizability from $k \geq 3$ onward (see the upper part of Table~\ref{solutions_table_HMS23}). Yet, this is highly sensitive to the utilities. If, for instance, we slightly change player 2's utility of $(b, e)$ from $2$ to $3$ (let's call it the HMS3 game), then strong rationalizability and iterated admissibility remain unchanged at all levels but now strong level-$k$ thinking (again with uniform initial beliefs) coincides with $k$-level strong rationalizability and is disjoint from $k$-iterated admissibility from $k \geq 3$ onward. See the lower part of Table~\ref{solutions_table_HMS23}. Again, this illustrates the sensitivity of predictions of strong level-$k$ thinking to small changes of utilities. It is due to the assumption of initial uniform beliefs.
\end{ex}

\begin{table}[h!]\caption{Solutions to HMS2 and HMS3 Games\label{solutions_table_HMS23}}
\tiny
\begin{center}
\begin{tabular}{||c||c|c||c|c||c|c||c|c||} \hline \hline HMS2 & \multicolumn{2}{c||}{Strong level-$k$} & \multicolumn{2}{c||}{$k$-level strong \& $k$-level} & \multicolumn{2}{c||}{$k$-iterated admissibility} & \multicolumn{2}{c||}{Backward} \\
 game & \multicolumn{2}{c||}{(uniform)} & \multicolumn{2}{c||}{backward rationalizability} & \multicolumn{2}{c||}{} & \multicolumn{2}{c||}{level-$k$ (uniform)} \\ \hline \hline
Level & Player 1 & Player 2 & Player 1 & Player 2 & Player 1 & Player 2 & Player 1 & Player 2 \\ \hline \hline
$1$ & $\{b\}$ & $\{d\}$ & $\{a, b, c\}$ & $\{d, e\}$ & $\{a, b\}$ & $\{d, e\}$ & $\{b\}$ & $\{d\}$ \\
$2$ & $\{a\}$ & $\{d\}$ & $\{a, c\}$ & $\{d, e\}$ & $\{a, b\}$ & $\{d\}$ & $\{a\}$ & $\{d\}$ \\
$3$ & $\{a\}$ & $\{d\}$ & $\{a, c\}$ & $\{e\}$ & $\{a\}$ & $\{d\}$ & $\{a\}$ & $\{d, e\}$ \\ 
$4$ & $\vdots$ & $\vdots$ & $\vdots$ & $\vdots$ & $\vdots$ & $\vdots$ & $\{a, c\}$ & $\{d, e\}$ \\ 
$5$ & $\vdots$ & $\vdots$ & $\vdots$ & $\vdots$ & $\vdots$ & $\vdots$ & $\{a, c\}$ & $\{e\}$ \\
$\vdots$ & $\vdots$ & $\vdots$ & $\vdots$ & $\vdots$ & $\vdots$ & $\vdots$ & $\vdots$ & $\vdots$ \\ \hline \hline
\multicolumn{9}{c}{} \\
\multicolumn{9}{c}{} \\
\hline \hline HMS3 & \multicolumn{2}{c||}{Strong level-$k$} & \multicolumn{2}{c||}{$k$-level strong \& $k$-level} & \multicolumn{2}{c||}{$k$-iterated admissibility} & \multicolumn{2}{c||}{Backward} \\
 game & \multicolumn{2}{c||}{(uniform)} & \multicolumn{2}{c||}{backward rationalizability} & \multicolumn{2}{c||}{} & \multicolumn{2}{c||}{level-$k$ (uniform)} \\ \hline \hline
Level & Player 1 & Player 2 & Player 1 & Player 2 & Player 1 & Player 2 & Player 1 & Player 2 \\ \hline \hline
$1$ & $\{b\}$ & $\{d\}$ & $\{a, b, c\}$ & $\{d, e\}$ & $\{a, b\}$ & $\{d, e\}$ & $\{b\}$ & $\{d, e\}$ \\
$2$ & $\{a, c\}$ & $\{d\}$ & $\{a, c\}$ & $\{d, e\}$ & $\{a, b\}$ & $\{d\}$ & $\{a, c\}$ & $\{d\}$ \\
$3$ & $\{a, c\}$ & $\{e\}$ & $\{a, c\}$ & $\{e\}$ & $\{a\}$ & $\{d\}$ & $\{a\}$ & $\{e\}$ \\ 
$4$ & $\vdots$ & $\vdots$ & $\vdots$ & $\vdots$ & $\vdots$ & $\vdots$ & $\{a, c\}$ & $\{d, e\}$ \\
$5$ & $\vdots$ & $\vdots$ & $\vdots$ & $\vdots$ & $\vdots$ & $\vdots$ & $\{a, c\}$ & $\{e\}$ \\
$\vdots$ & $\vdots$ & $\vdots$ & $\vdots$ & $\vdots$ & $\vdots$ & $\vdots$ & $\vdots$ & $\vdots$ \\ \hline \hline
\multicolumn{9}{c}{} \\
\multicolumn{9}{c}{} \\
\end{tabular}
\end{center}
\normalsize
\end{table}

Although iterated admissibility captures cautiousness, it is a solution concept for the normal form. While Kohlberg and Mertens (1986) argue that the (associated) game in normal-form should be sufficient for solutions to games, there is evidence that behaviorally the extensive-form versus normal-form representation of the games makes a difference (e.g., Cooper and Van Huyck, 2003). It would therefore be desirable to also have a strong rationalizability concept that captures cautiousness. Heifetz, Meier, and Schipper (2021) put forward the following definition of prudent rationalizability.\footnote{It has been applied to partially identify cautious level-$k$ reasoning in experiments by Li and Schipper (2020). It also been applied to games with unawareness including disclosure games (Heifetz, Meier, and Schipper, 2021; see Li and Schipper, 2019, for experiments), electoral campaigning (Schipper and Woo, 2019) and, with additional belief restrictions, to screening problems (Francetich and Schipper, 2022). In Schipper and Woo (2019), the levels of reasoning embodied in prudent rationalizability have been used to model the political reasoning capabilities of voters.}

\begin{defin}[Prudent Rationalizability] For any $i \in N$, let
$\hat{R}_{i}^{0}=S_{i}.$ For $k\geq 1$, define inductively
\begin{equation*}
\begin{array}{rcl}
\hat{B}_{i}^{k}\bigskip  & =\bigskip  & \left\{ \bar{\beta}_{i}\in \bar{B}_{i}:
\begin{array}{l}
\mbox{For every information set }I_{i}, \mbox{ if there exists some
profile } \\
s_{-i }\in \hat{R}_{-i}^{k-1} \mbox{ of the other players' strategies such that }s_{-i} \\
\mbox{reaches }I_{i}, \mbox{ then the support of }\beta_{i}\left( I_{i}\right)
\mbox{is the set of strategy} \\\mbox{profiles }s_{-i}\in \hat{R}_{-i}^{k-1}\mbox{ that reach }I_{i}.
\end{array}
\right\}
\end{array}%
\end{equation*}%
\begin{equation*}
\begin{array}{rcl}
\hat{R}_{i}^{k} & = & \left\{ s_{i}\in \hat{R}_{i}^{k-1}:\begin{array}{l}
\mbox{There exists }\beta_{i}\in \hat{B}_{i}^{k}\mbox{ such that for all }
I_{i}\in \mathcal{I}_{i} \\
\mbox{player } i \mbox{ with strategy } s_i \mbox{ is rational at }I_{i}.
\end{array}
\right\}
\end{array}
\end{equation*}
The set of prudent rationalizable strategies of player $i$ is
\begin{equation*}
\hat{R}_{i}^{\infty }=\bigcap_{k=1}^{\infty }\hat{R}_{i}^{k}.
\end{equation*}
\end{defin}

Note that this solution concept features non-nested sets of beliefs but nested sets of strategies. It is clear that a full-support belief on a smaller opponents' strategy subspace cannot be an element of the full-support beliefs on larger opponents' strategy subspaces. Thus, the set of $k$-level prudent belief systems cannot be a subset of the set of $k-1$ level prudent belief systems. Heifetz, Meier, and Schipper (2021) show that it is nonempty for any finite game in extensive form (including games in extensive form with unawareness).

Meier and Schipper (2023) show that prudent rationalizability is level-by-level strategy equivalent to iterated admissibility in the associated normal form including games with unawareness (see Shimoji and Watson, 1998, and Brandenburger and Friedenberg, 2011, for related results). Thus, Example~\ref{HMS_example} demonstrates already how strong level-$k$ thinking with level-$1$ uniform belief systems differs from prudent rationalizability.

\subsection{Strong Level-$k$ Thinking versus Backward Rationalizability}

For the following subsections, we assume that the games in extensive form are such that there is a precedence relation on the information sets of the game. Consequently, for any information sets $I, I' \in \bigcup_{i \in N} \mathcal{I}_i$, we write $I \preceq I'$ if information set $I$ precedes information set $I'$. 

By the logic of backward induction or subgame perfection, players just care about behavior in the continuation game. Thus, continuation strategies become important. For any players $i, j \in N$, strategy $s_i \in S_i$, and information set $I \in \mathcal{I}_j$, denote by $s_{i \mid I}$ the corresponding strategy in the continuation game starting at information set $I$ if player $i$ has an information set in the continuation game starting at $I$. That is, if player $i$ has an information set in the continuation game starting at $I$, then $s_{i \mid I}(I') := s_i(I')$ for all $I' \in \mathcal{I}_i$ with $I \preceq I'$. Similarly, for any nonempty subset of strategies $X_i \subseteq S_i$ and information set $I \in \mathcal{I}_j$, denote by $X_{i \mid I}$ the subset of strategies of player $i$ in the continuation game starting at information set $I$, i.e., $X_{i \mid I} := \{s_{i \mid I} : s_i \in X_i\}$. Note that $X_{i \mid I}$ might be empty if player $i$ has no information set in the continuation game starting at information set $I$. 

Although with the backward induction logic, players just care about continuation strategies, they form beliefs over strategies. To this end, we let $[X_{i \mid I}] : = \{s_i \in S_i : \mbox{If } X_{i \mid I} \neq \emptyset, \mbox{ then } s_{i \mid I } \in X_{i \mid I}\}$. This is the set of player $i$'s strategies whose continuation strategies are in $X_{i \mid I}$. Intuitively, when a player believes in $[X_{i \mid I}]$, she does not care how $I$ has been reached with player $i$'s strategies. She just cares about player $i$'s continuation strategies in $X_{i \mid I}$. Note that if $X_{i \mid I} = \emptyset$, then $[X_{i \mid I}] = S_i$. That is, $[X_{i \mid I}]$ is always nonempty by definition. Lastly, we let $[X_{-i \mid I}] := \times_{j \in N \setminus \{i\}} [X_{j \mid I}]$. Note that we may have $X_{-i} \subseteq [X_{-i \mid I}]$. For instance, not all strategies in a game may reach information set $I$. In such a case, we may have $S_{-i}(I) \neq [S_{-i \mid I}(I)]$. The l.h.s. denotes the set of strategy profiles of player $i$'s opponents that reach $I$ while the r.h.s. denotes the set of all strategy profiles of player $i$'s opponents whose continuation strategy profiles from information set $I$ onward coincide with continuation strategy profiles of strategy profiles of player $i$'s opponents that reach $I$. For instance, in the centipede game (Figure~\ref{centipede}), the second level strong rationalizable strategies of player 1 reaching the first information set of player 2 are $\{(C_1, o_1)\}$ but the strategies of player 1 coinciding with the continuation strategies of strategies of player 1 who reach this information set are $\{(C_1, o_1), (O_1, o_1)\}$. Latter type of strategies are relevant for backward rationalizability because, as we will see shortly, at every information set the player does not care how it is reached. 

The following definition is due to Penta (2010) and Perea (2014). In fact, the present version is a slight generalization because Penta (2010) just states it for multi-stage games with observable actions and Perea (2014) states it for complete information games. Meier and Perea (2023) define a version for games with time periods.    

\begin{defin}[Backward Rationalizability] Define inductively for every player $i \in N$, 
\begin{eqnarray*} \vec{B}_i^1 & = & \bar{B}_i \\
\vec{R}_i^1 & = & \left\{s_i \in S_i : \begin{array}{l} \mbox{There exists } \bar{\beta}_i \in \vec{B}_i^1 \mbox{ such that for every information} \\ \mbox{set } I_i \in \mathcal{I}_i, s_{i \mid I_i} \mbox{ is rational at } I_i \mbox{ given } \bar{\beta}_i(I_i). \end{array}\right\} \\
& \vdots & \\
\vec{B}_i^k & = & \left\{ \bar{\beta}_i \in \vec{B}_i^{k-1} : \begin{array}{l} \beta_i(I_i)\left(\left[\vec{R}_{-i \mid I_i}^{k-1}\right]\right) = 1 \mbox{ for every information set } I_i \in \mathcal{I}_i. \end{array} \right\} \\
\vec{R}_i^k & = & \left\{s_i \in S_i : \begin{array}{l} \mbox{There exists } \bar{\beta}_i \in \vec{B}_i^k \mbox{ with which for every information} \\ \mbox{set } I_i \in \mathcal{I}_i, s_{i \mid I_i} \mbox{ is rational at } I_i \mbox{ given } \bar{\beta}_i(I_i).\end{array} \right\} 
\end{eqnarray*} The set of backward rationalizable strategies of player $i$ is $$\vec{R}_i^{\infty}  = \bigcap_{k = 1}^{\infty} \vec{R}_i^k.$$ 
\end{defin}

A couple of remarks are in order: First, the major difference between backward rationalizability and strong rationalizability is that at an information set players believe at level $k$ that the opponents play $k-1$ backward rationalizable continuation strategies from there onward no matter whether the information set could have been reached with such $k-1$ backward rationalizable strategies or not. Since this holds for every continuation game, it is very much in spirit of subgame perfection, or more generally, ``continuation-game perfection.'' In fact, it may have been more appropriately dubbed ``continuation-game rationalizability.'' However, Cantonini and Penta (2022) show (for multi-stage games) that backward rationalizability is equivalent to a backwards induction rationalizability procedure starting from a subgame of lowest rank and moving backward to successively larger subgames. So in this sense, the name backward rationalizability is justified. Cantonini and Penta (2022) also characterize further properties of the solution concept such as order-independence and continuation-game consistency. Perea (2014) shows equivalence to backward dominance for complete information games. More importantly, he characterizes it by common belief in opponents behaving rationally in future. Battigalli and De Vito (2021) epistemically characterize backward rationalizability in multi-stage games with observable actions in a richer epistemic framework, in which they are able to distinguish between planed play and actual play. 

How does $k$-level backward rationalizability compare to strong level-$k$ thinking? One defining feature of strong level-$k$ thinking is forward induction, while this is absent in backward rationalizability. This difference causes strategies of strong level-$k$ thinking (with uniform initial belief systems) to differ from backward rationalizability for instance in the Reny game (Figure~\ref{Reny}). At level-$3$, player 2 would choose continue at her first information sets with strong level-$k$ thinking while moving out with the backward rationalizability (see Table~\ref{solutions_table}). This is due to backward rationalizability following the backward induction logic. Since player 1 moves out with both solution concepts, the solution concepts do not yield different outcomes in this game. However, strong level-$k$ thinking (with uniform initial belief systems) is a refinement of backward rationalizable outcomes in for instance the HMS2 game (Figure~\ref{HMS2}). There, strong level-$k$ thinking yields outcome $a$ from level-$2$ upward while backward rationalizability allows for strategy $c$ in addition to $a$ (see Table~\ref{solutions_table_HMS23}). Similarly, strong level-$k$ thinking (with uniform initial belief systems) is a outcome refinement of backward rationalizability in the Battle-of-the-sexes games with an outside option I and II (see Table~\ref{BoS_table}).

\subsection{Strong Level-$k$ versus Backward Level-$k$ Thinking} 

Ho and Su (2013) study centipede games with a notion of ``dynamic level-k.'' Here the dynamics w.r.t. levels is \emph{between} repeated play of centipede stage-games, \emph{not within} the centipede game. However, within the centipede game, they use what we could dub ``subgame level-k''. That is, they assume first level beliefs in random choice at every history of the game. For any higher level, they require that the $k$-level strategy is a best response to the $(k - 1)$-level strategies of opponents in \emph{every} subgame no matter whether the subgame is reached or not. Kawagoe and Takizawa (2012) also present an experimental analysis of centipede games using level-$k$ best responses at every history of the game like Ho and Su (2013) (see also Stahl and Haruvy (2008) for two-stage games). However, they explicitly assume additional noise which makes any subgame reachable. In the following, we define formally a notion of `backward level-$k$' thinking, i.e., level-$k$ thinking at \emph{every} subgame. More generally, since we allow for moves of nature, simultaneous moves of players, and non-singleton information sets, it should be more appropriately understood as a notion of ``continuation-game level-$k$'' thinking. 

\begin{defin}[Backward Level-$k$ Thinking] Given a system of first-level beliefs, $\bar{\beta}^1 = (\bar{\beta}_i^1)_{i \in N}$ with $\bar{\beta}_i^1 \in \bar{B}_i$, define inductively for every player $i \in N$, 
\begin{eqnarray*} \vec{B}_i^1(\bar{\beta}^1) & = & \{\bar{\beta}_i^1\} \\
\vec{L}_i^1(\bar{\beta}^1) & = & \left\{s_i \in S_i : \begin{array}{l} \mbox{For every information set } I_i \in \mathcal{I}_i, \\
s_{i \mid I_i} \mbox{ is rational at } I_i \mbox{ given } \bar{\beta}_i^1(I_i). \end{array}\right\} \\
& \vdots & \\
\vec{B}_i^k(\bar{\beta}^1) & = & \left\{ \bar{\beta}_i \in \bar{B}_i : \begin{array}{l} \bar{\beta}_i(I_i)\left(\left[L_{-i \mid I_i}^{k-1}(\bar{\beta}^1_i)\right]\right) = 1 \mbox{ for every information set } I_i \in \mathcal{I}_i. \end{array} \right\} \\
\vec{L}_i^k(\bar{\beta}^1) & = & \left\{s_i \in S_i : \begin{array}{l} \mbox{There exists } \bar{\beta}_i \in \vec{B}_i^k(\bar{\beta}^1) \mbox{ with which for every information} \\ \mbox{set } I_i \in \mathcal{I}_i, s_{i \mid I_i} \mbox{ is rational at } I_i \mbox{ given } \bar{\beta}_i(I_i).\end{array} \right\} 
\end{eqnarray*}
\end{defin}

At level $1$, each player has beliefs over opponents' strategies given by the belief system $\bar{\beta}^1$. Typically, the literature assumes some level-$0$ type that mixes uniformly at every information set. However, since these types are just fictitious, we model them as what they are, namely beliefs of level-$1$ types. Level-$1$ types best respond to their beliefs at every continuation game. At any level $k$ and information set of a player, the player forms beliefs about $(k-1)$-level continuation strategies of opponents. Of course, it could be that no level-$(k-1)$ strategy profile of opponents can reach the information set. In that case, the player nevertheless believes that opponents play continuation strategies consistent with level-$(k-1)$ strategies. That is, the player assumes that from now on, opponents play level-$(k-1)$ strategies. If opponents have no action in the continuation game, then the player is free to believe anything. 

The purpose of introducing backward rationalizability before backward level-$k$ thinking is to make clear that backward level-$k$ thinking can be interpreted as a form of $k$-level backward rationalizability with a restriction on first-order belief systems. In particular, Ho and Su (2013) assumed uniform first-order beliefs at every information set. So from a conceptual point of view, the relationship between backward level-$k$ thinking and level-$k$ backward rationalizability is analogous to the relationship between strong level-$k$ thinking and level-$k$ strong rationalizability. Of course, strategies and outcomes may differ. 

Backward level-$k$ thinking is a solution concept different from strong level-$k$ thinking as latter features forward induction while former does not. However, somewhat perplexingly at a first glance in some games backward level-$k$ thinking with uniform initial belief systems can yield forward induction outcomes of let's say strong rationalizability or iterated admissibility while strong level-$k$ thinking with uniform initial belief systems does not. Consider for instance the Battle-of-the-sexes with an outside option game I (see Figure~\ref{BoS1}). At level-$3$, backward level-$k$ thinking yields uniquely the strategy profile $((In, B), B)$ (see Table~\ref{BoS_table}, upper part), which is the ``classic'' forward induction outcome of strong rationalizability and iterated admissibility in these games. However, closer inspection shows that backward level-$k$ thinking yields this forward induction outcome without forward induction reasoning behind it. It is simply due to player 1 best responding to player 2 best response to player 1's continuation strategy in the Battle-of-the-sexes game. Latter is a best response in this subgame to player 1's uniform belief over player 2's actions. In contrast, strong level-$k$ with uniform initial belief systems yields the set of strategy profiles $\{((Out, S), S), ((Out, B), S)\}$ and thus a outcome different from backward level-$k$ thinking and typically not associated with forward induction. This outcome is due to player 1 best responding to his uniform belief about player 2's actions and choosing out. Since the Battle-of-the-sexes subgame is not reached, player 2 resorts to her first-level beliefs and assumes that player 1 would behave uniformly in the Battle-of-the-sexes subgame. Consequently, she keeps choosing action $S$ at higher levels. More generally, the example demonstrates that backward level-$k$ thinking and strong level-$k$ thinking are distinct solution concepts since in this game they lead to different outcomes from level-$3$ upward. However, we observe the following: 

\begin{prop}\label{first-level_equivalence} For any initial profile of full support belief systems $\bar{\beta}^1$, the outcomes of strong level-$1$ strategies coincide with outcomes of backward level-$1$ strategies. 
\end{prop}

As the proof of the proposition in the appendix notes, the observation does not extent to strategies. That is, strong level-$1$ strategies may differ from backward level-$1$ strategies since we can have the case that $s_i$ does not reach some information set $I_i$. In such a case $s_i$ is trivially rational at $I_i$ but $s_{i \mid I_i}$ may not be rational at $I_i$. An example is again the Battle-of-the-sexes game with an outside option I of Figure~\ref{BoS1}. As we noted before, the set of strong level-$1$ strategies of player 1 is $\{(Out, S), (Out, B)\}$ while only $(Out, B)$ is the unique backward level-$1$ strategy. It does not mean though that backward level-$k$ thinking is a refinement of strong level-$k$ thinking. We noted this already in the same game where at level 3 the two solution concepts yield distinct outcomes. 

As we have seen already with normal-form level-$k$ and strong level-$k$, also backward level-$k$ can lead to cyclic choices across levels. Examples are the Battle-of-the-sexes games with an outside option II and III (Figures~\ref{BoS2} and~\ref{BoS3}), where backward level-$k$ leads to a cycle of alternating strategies listed in Tables~\ref{BoS_table} and~\ref{BoSIII_solution}. Again, such a cycle creates difficulties for an epistemic characterization but corresponds nicely to Morgenstern's observation cited in the Introduction that in some games level-by-level reasoning about opponents does not lead to a solution.

\subsection{Strong Level-$k$ Thinking versus Backward Induction}

Arguably the most commonly used solution concept to games in extensive form with perfect information and sequential moves is subgame-perfect equilibrium solved by backward induction. We can consider backward induction as an iterated process ``level-by-level'', where now a ``level'' refers to the rank of a subgame rather than a level of belief. Consider for simplicity a finite game in extensive form with perfect information, sequential moves, and a finite horizon. Call the rank of a subgame the maximal number of nodes to reach a terminal node. Backward induction is now defined as follows: At level $1$, consider all subgames of rank 1. Select a utility maximizing action of the player that moves in this subgame. Replace the subgame with a terminal node ascribing this utility to this newly created terminal node. Assume we have defined the procedure at level $k-1$. Then at level $k$, consider each subgame of rank $k$ (in the original game in extensive form). Select a utility maximizing action of the player that moves in this subgame considering the utilities obtained from the procedure at level $k-1$. Replace the subgame with a terminal node ascribing this utility to this newly created terminal node. Do this with all subgames of rank $k$. Since the finite game has a finite horizon, the procedure stops after some finite number of levels.

Although the notion of level in backward induction is very different from the notion of level in various notions of level-$k$, we can compare strong level-$k$ thinking with backward induction up to level $k$. The following examples demonstrate that $k$-level backward induction and strong level-$k$ thinking are quite distinct solution concepts. One of the most prominent game displaying the awkward backward induction logic transparently is the centipede game (Rosenthal, 1981). In the centipede game, strong level-$k$ thinking may refine level-$k$ backward induction at some levels but is refined by level-$k$ backward induction at some other levels.

\begin{ex}[Centipede game]\label{centipede_example} Consider a short version of the centipede game depicted in Figure~\ref{centipede}.
\begin{figure}[h!]\caption{Centipede game\label{centipede}}
	\begin{center}
    \includegraphics[scale=0.4]{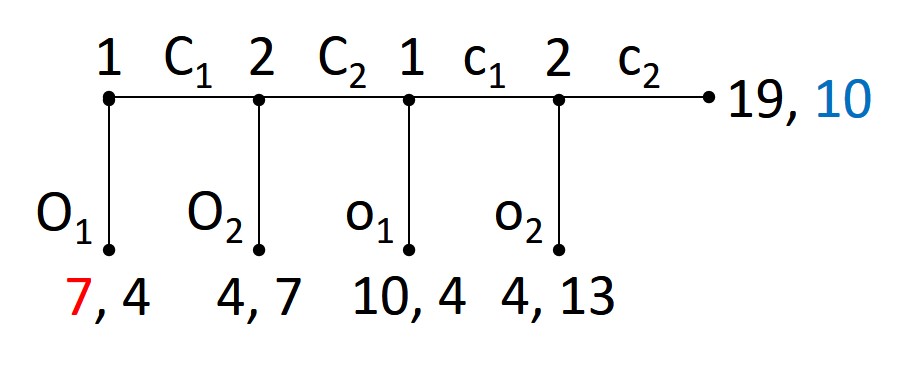} 	\includegraphics[scale=0.4]{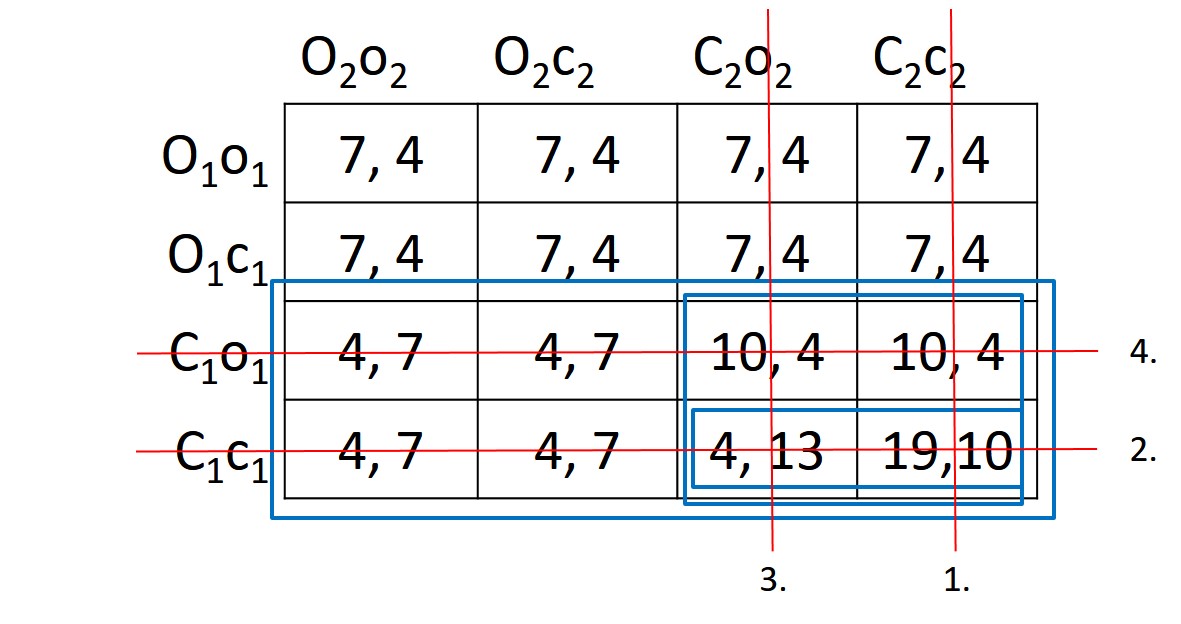}
    \end{center}
\end{figure}
The set of strong level-$k$ strategies (with uniform initial belief systems) and backward induction strategies for every level $k$ are detailed in Table~\ref{solutions_table}. We note that at level-$1$, strong level-$k$ thinking (with uniform initial belief system) refines the level-$1$ backward induction strategies and outcomes. This is not surprising as we reap the full power of the assumption on the particular uniform``level-$0$'' behavior while only having the refinement power of backward induction for the game of rank 1, the ``tail'' and ``last leg'' of the centipede. At level $2$, strong level-$k$ thinking still refines level-$2$ backward induction but with strategies that differ starkly from strong level-$1$ thinking as players now take strong level-$1$ strategies of the opponent into account. At level-$3$, strong level-$k$ thinking still refines level-$3$ backward induction outcomes but the strategy of player 2 is now inconsistent with backward induction. The reason is that player 2 does not expect that his second information set is reached. Hence any action at this information set is rational. In contrast, if player 2 follows backward induction and reaches her second information set, she ``shrugs her shoulders'' and ignorantly continues to do backward induction as if nothing had happened. As from level-$4$ though, backward induction refines strong level-$k$ strategies (for the same reason that produced the difference at level 3). The outcomes are the same though. To summarize, for levels $k \leq 3$, strong level-$k$ thinking refines level-$k$ backward induction in terms of outcomes. Yet, for levels $k \geq 4$, strong level-$k$ thinking is refined by level-$k$ backward induction in terms of strategies but not outcomes. This changing pattern of the relationship between strong level-$k$ thinking and level-$k$ backward induction highlights the fact that they are conceptually quite different solution concepts although both can be understood as solutions concepts employing some form of inductive elimination of strategies. We note that strong level-$k$ thinking does not display the often criticized ``stubborn belief'' in the opponent playing backward induction when an information set is reached that is not on the outcome path. Not surprisingly, $k$-level backward rationalizability and backward level-$k$ yields the same strategies as backward induction from level-4 upward since all these solution concepts capture the idea of subgame perfection. 
\end{ex}

\section{Revisiting Prior Experiments\label{experiments}}

There is a large literature on testing solution concepts to games in extensive form in experimental game theory. Given our observations with regard to strong level-$k$ reasoning, normal-form level-$k$ reasoning, and $k$-level strong rationalizability, data sets on games in which forward induction plays a role are of special interest to us. We are very grateful to authors of some previously published experiments for providing us with their data sets. In this section, we report as a proof of concept on a simple reanalysis of those extant data sets attempting to glean different aspects of strategic sophistication.

\subsection{The Role of Forward Induction Given the Level of Thinking and Uniform First-Level Beliefs\label{Cooper_section}}

\begin{figure}[h!]\caption{Game used by Cooper et al. (1993)\label{BoS_Cooper_300}}
	\begin{center}
		\includegraphics[scale=0.4]{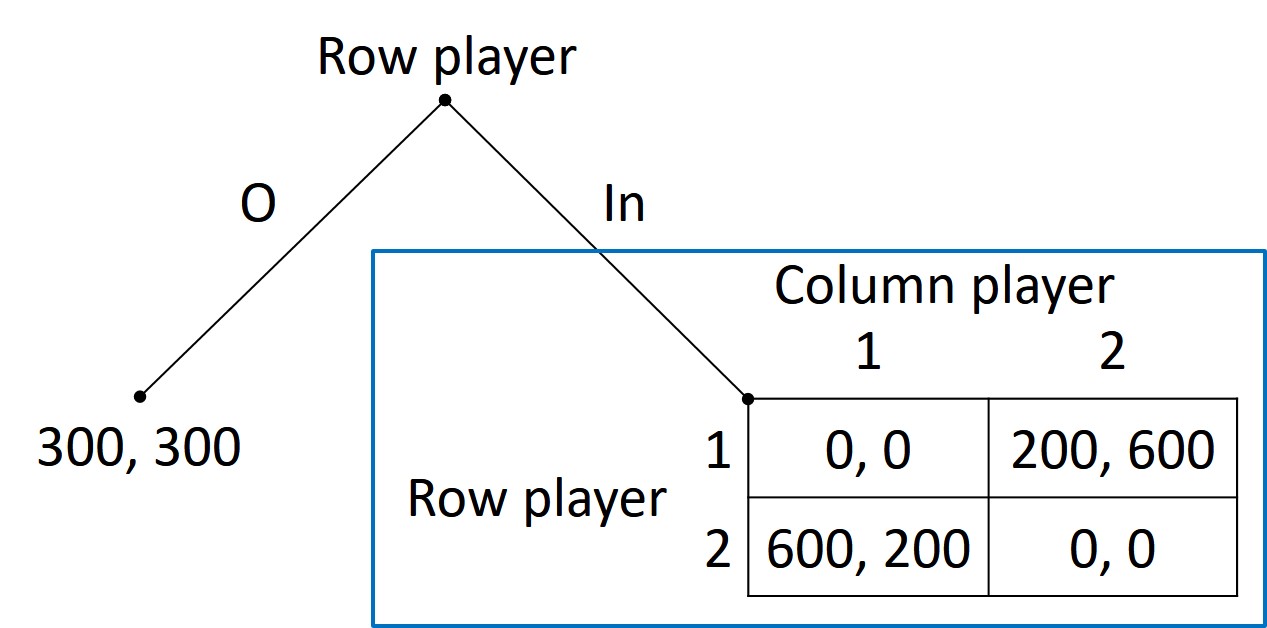} 	
	\end{center}
\end{figure}

Cooper et al. (1993) conducted experiments on the Battle-of-the-Sexes game with outside options. One of the games used in their experiments is depicted in Figure~\ref{BoS_Cooper_300}. Although its proper subgame is phrased as an ``anti-coordination'' game, it is seen easily to be equivalent to a Battle-of-the-Sexes game just by renaming the actions of one player.

Table~\ref{BoS_Cooper_300_solutions} presents level-by-level three solutions: strong level-$k$ with uniform initial belief systems, normal-form level-$k$ with uniform initial beliefs, and $k$-level strong rationalizability.
\begin{table}[h!]\caption{Solutions to the game used by Cooper et al. (1993) \label{BoS_Cooper_300_solutions}}
	\footnotesize
	\centering
\begin{tabular}{||c||c|c||c|c||c|c||}\hline \hline & \multicolumn{2}{c||}{Strong level-$k$} & \multicolumn{2}{c||}{Normal-form level-$k$} & \multicolumn{2}{c||}{$k$-level strong} \\
& \multicolumn{2}{c||}{(uniform)} & \multicolumn{2}{c||}{(uniform)} & \multicolumn{2}{c||}{rationalizability} \\ \hline \hline
Level & Row player  & Column pl. & Row player & Column pl. & Row player & Column pl. \\ \hline \hline
$1$ & \{(In, 2),(O, *)\} & 2 & \{(In, 2),(O, *)\} & 2 & \{(In, 2),(O, *)\} & \{1, 2\} \\
$2$ & (O, *)  & 1 & (O, *) & 1 & \{(In, 2),(O, *)\} & 1 \\
$3$ & (In, 2) & 1 & (In, 2) & \{1, 2\} & (In, 2) & 1 \\
$4$ & (In, 2) & 1 & \{(In, 2),(O, *)\} & 1 & (In, 2) & 1 \\
\vdots & \vdots & \vdots & \vdots & \vdots & \vdots & \vdots \\
\hline \hline
\end{tabular}
\end{table} We observe that in this game, strong level-$k$ reasoning is equivalent to $k$-level strong rationalizability from the third level onward. It refines $k$-level strong rationalizability at the first two levels. More importantly, strong level-$k$ thinking is a strategy refinement of normal-form level-$k$ reasoning from level 3 onward and an outcome refinement of normal-form level-$k$ reasoning from level 4 onward. Both, strong level-$k$ reasoning and normal-form level-$k$ reasoning feature levels of reasoning. Yet, strong level-$k$ reasoning also features a second dimension of strategic sophistication, namely the ability to update beliefs about opponent's future behavior given the opponent's past behavior. Games like the present one, in which strong level-$k$ thinking strictly refines normal-form level-$k$ thinking, provide us with an opportunity to assess the importance of the forward-induction ability beyond just level-$k$ thinking. This becomes apparent at level 3. Under normal-form level-$k$ reasoning for $k = 3$, the column player is indifferent between actions 1 and 2 thinking that the row player chooses Out anyway. In contrast, when the column player's information set is reached and she gets to play, she now knows under strong level-$k$ reasoning with $k = 3$ that the row player cannot be an strong level-$2$ reasoner because such a row player would move Out. At strong level-3, the only way for the column player to make sense of the row player's action to move In is to attribute level-1 to the row player. Here we see the ``best rationalizability principle'' embodied in strong level-$k$ thinking at work. Rather than believing that the row player is irrational by choosing In, the column player attributes the highest level of rationality consistent with reaching the subgame to the row player (and below her own level-$3$), which is strong level-$1$. This makes her realize that the row player plans to play 2 since he moved In already. Thus, she best responds with taking action 1.

Participants played the game for 22 periods and alternated between the row and column player positions. Players were anonymously re-matched within each of the 21 cohorts. Results of the first 11 periods differed significantly from the last 11 periods and Cooper et al. (1993) reported only on the last 11 periods. Other treatments of the experiment involved variants of the Battle-of-the-Sexes games with outside options. We focus on the game of Figure~\ref{BoS_Cooper_300} because strong level-$k$ is a \emph{strict} refinement of normal-form level-$k$ reasoning.\footnote{For instance, Cooper et al. (1993) also report on a treatment involving a game similar to Example~\ref{BoSIII_example}. However, in this game, strong level-$k$ coincides with normal-form level-$k$ and strong rationalizability provides only a coarse solution. Therefore, we do not think we can learn much for our purposes from that treatment w.r.t. the solution concepts discussed here and omit a reanalysis.} See Cooper et al. (1993) for further details of the experimental design.\footnote{Although we received the data from Cooper et al. (1993), we were not able to fully comprehend them yet given that they were collected more than 30 years ago. Thus, our analysis makes use of the frequencies reported in Cooper et al. (1993, Table 4).}

Table~\ref{BoS_Cooper_300_results} describes the percentage of choices from the last 11 rounds consistent with the various solution concepts.
\begin{table}[h!]\caption{Choices in Cooper et al. (1993) consistent with solutions\label{BoS_Cooper_300_results}}
	\footnotesize
	\centering
\begin{tabular}{||c||c|c||c|c||c|c||}\hline \hline & \multicolumn{2}{c||}{Strong level-$k$} & \multicolumn{2}{c||}{Normal-form level-$k$} & \multicolumn{2}{c||}{$k$-level strong} \\
& \multicolumn{2}{c||}{(uniform)} & \multicolumn{2}{c||}{(uniform)} & \multicolumn{2}{c||}{rationalizability} \\ \hline \hline
Level & Row player  & Column pl. & Row player & Column pl. & Row player & Column pl. \\ \hline \hline
$1$ & 98\% & 8\% & 98\% & 8\% & 98\% & 100\% \\
$2$ & 20\%  & 92\% & 20\% & 92\% & 98\% & 92\% \\
$3$ & 78\% & 92\% & 78\% & 100\% & 78\% & 92\% \\
$4$ & 78\% & 92\% & 98\% & 92\% & 78\% & 92\% \\
\hline \hline
\end{tabular}
\end{table}
First, we observe that a large percentage of choices are consistent with all three solution concepts. Second and more importantly, we note that apparently the second dimension of strategic sophistication, forward-induction, is just missing in 8\% of the column players \emph{at comparable levels of reasoning} (i.e., at level-$3$, the relevant level for forward induction) and \emph{comparable assumptions on first-level beliefs/level-0 behavior} (i.e., uniform). Normal-form level-$k$ with uniform initial beliefs trivially fits 100\% of the data on the column player for $k = 3$ while strong level-$k$ with uniform initial belief systems fits 92\% of the data at $k = 3$. 

The exercise also offers a glimpse of how strong level-$k$ thinking might be used in experimental game theory. The goal is not so much in winning a horse race among solution concepts in a fitting exercise. Rather, by comparing different solution concepts who differ just in one particular feature of strategic sophistication from each other but are otherwise comparable, we might learn about the prevalence of this feature without interference by other varying features, which is very much in the spirit of comparative statics. When comparing the fit of strong level-$k$ with uniform initial belief systems and normal-form level-$k$ with uniform initial beliefs in the experiment, we learn about the prevalence of forward induction \emph{given} comparable levels of thinking and comparable assumptions on initial beliefs.

\subsection{The Role of Uniform Level-$1$ Beliefs Given Forward Induction and the Level of Reasoning\label{BN_section}}

\begin{figure}[h!]\caption{Game used by Balkenborg and Nagel (2016)\label{BN2016}}
	\centering
	\includegraphics[scale=0.4]{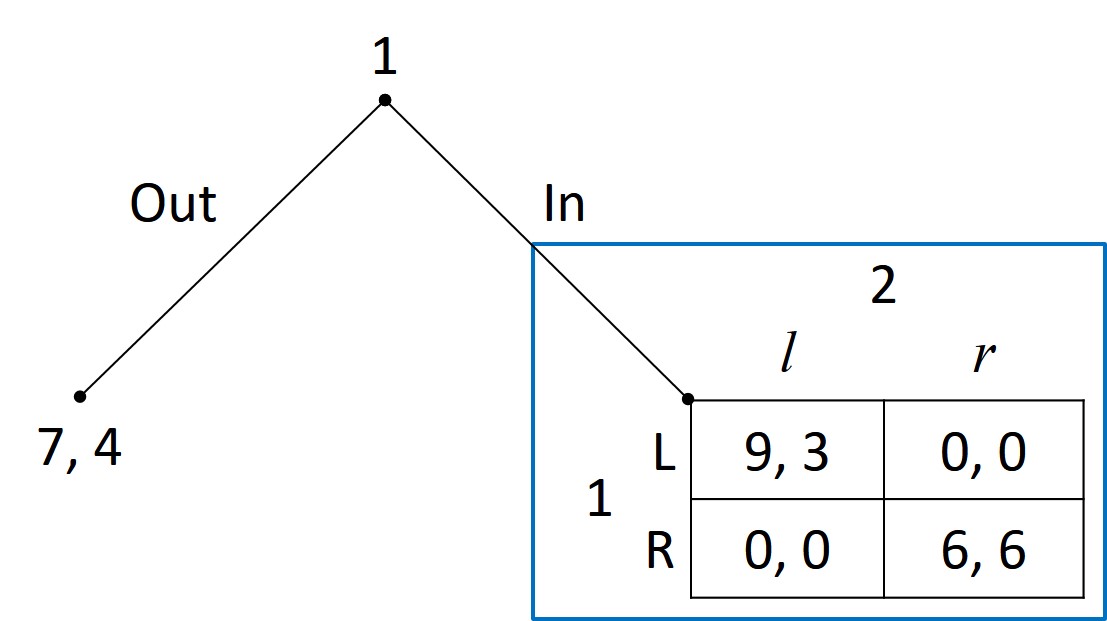}
\end{figure}

Balkenborg and Nagel (2016) study a variant of the Battle-of-the-sexes game with an outside option in which nature moves first and selects between the outside option game or the Battle-of-the-sexes game without the outside option. We focus here on the subgame consisting of the Battle-of-the-sexes game with the outside option as depicted in Figure~\ref{BN2016}.\footnote{In their terminology, it is the ``left game''.} While such a focus on a subgame only is generally problematic when studying solution concepts with forward induction, because beliefs of players might be affected by what happened before the subgame, we do not think that it does affect our analysis of this particular game in a relevant way.

The predictions of strong level-$k$ thinking (with uniform initial belief systems) and $k$-level strong rationalizability are given in Table~\ref{BN_table}. It is well-known that the outcome of strong rationalizability is equivalent in this game to the prediction by iterated admissibility and strategic stability. Balkenborg and Nagel (2016) refer to this outcome simply as the forward induction outcome. Their interest is on testing it against Harsanyi-Selten equilibrium selection based on risk-dominance, the focal point, and strong backward induction. In particular, since the action profile $(R, r)$ is both risk-dominant and focal due to symmetry of payoffs, $(6, 6)$, backward induction suggests that player 1 chooses out and guarantees himself the larger payoff of 7. This prediction coincides with strong level-$k$ thinking with uniform level-$0$ belief systems in this game for $k \geq 1$. In fact, already Balkenborg and Nagel (2016, p. 398) note that Out is player 1's best response to the uniform belief over player 2's actions in the Battle-of-the-sexes subgame. The fact that strong level-$k$ thinking with uniform level-$0$ belief systems differs from $k$-level strong rationalizability in this game allows us to study the effect of level-$0$ uniform beliefs given the ability to do forward induction and given comparable levels of thinking.
\begin{table}[h!]\caption{Solutions to the game of Balkenborg and Nagel (2016)\label{BN_table}}
\scriptsize
\centering
\begin{tabular}{||c||c|c||c|c||c|c||} \hline \hline & \multicolumn{2}{c||}{Strong level-$k$} & \multicolumn{2}{c||}{Normal-form level-$k$} & \multicolumn{2}{c||}{$k$-level strong}  \\
 game & \multicolumn{2}{c||}{(uniform)} & \multicolumn{2}{c||}{(uniform)} &  \multicolumn{2}{c||}{rationalizability} \\ \hline \hline
Level & Player 1 & Player 2 & Player 1 & Player 2 & Player 1 & Player 2 \\ \hline \hline
$1$ & $(Out, *)$ & $r$ & $(Out, *)$ & $r$ & $\{(Out, *), (In, L)\}$ & $\{l, r\}$ \\
$2$ & $(Out, *)$ & $r$ & $(Out, *)$ & $\{l, r\}$ & $\{(Out, *), (In, L)\}$ & $l$ \\
$3$ & $\vdots$ & $\vdots$ & $\{(Out, *), (In, L)\}$ & $\{l, r\}$ & $\{(In, L)\}$ & $\vdots$ \\
$4$ & $\vdots$ & $\vdots$ & $\vdots$ & $\vdots$ & $\vdots$ & $\vdots$ \\ \hline \hline
\end{tabular}
\end{table}

Note that normal-form level-$k$ reasoning with uniform initial beliefs has ``no bite'' in this game. This is because at level 3, player 1 can have arbitrary beliefs about both level-2 best responses of player 2. This makes both $(Out, *)$ and $(In, L)$ consistent with normal-form level-$k$ thinking for $k \geq 3$. It just underlines the fact that normal-form solution concepts are not always useful for studying games in extensive form. That's why in what follows we focus on strong level-$k$ thinking and $k$-level strong rationalizability.

In their experiments, 154 students participated in 13 independent sessions. In each session, the game was played sequentially for 50 rounds which was followed by one round of play using the strategy method (Selten, 1967). Participants were randomly rematched after each round but maintained their player role throughout the experiment. Sessions differed by the information feedback but results did not differ so that data of the various sessions have been pooled together.
\begin{table}[h!]\caption{Choices in Balkenborg and Nagel (2016)\label{BN_choices_table}}
\footnotesize
\centering
\begin{tabular}{||c||c|c||c|c||} \hline \hline & \multicolumn{2}{c||}{Strong level-$k$} & \multicolumn{2}{c||}{$k$-level strong}  \\ & \multicolumn{2}{c||}{(uniform)} & \multicolumn{2}{c||}{rationalizability} \\ \hline \hline
Level & Player 1 & Player 2 & Player 1 & Player 2 \\ \hline \hline
$1$ & $88\% $ & $43\% $ & $90.2\%$ & $100\%$ \\
$2$ & $88\% $ & $43\% $ & $90.2\%$ & $57\% $ \\
$3$ & $88\% $ & $43\% $ & $2\% $& $57\% $ \\ \hline \hline
\end{tabular}
\normalsize
\end{table}

We classify individual choices from all periods in Table~\ref{BN_choices_table}. Our exercise shows that for the Balkenborg and Nagel (2016) dataset, 88\% of the player 1 chooses Out, which is consistent with the prediction of strong level-$k$ thinking but not with $k$-level strong rationalizability from level-$3$ onward. Since the most substantial difference between $k$-level strong rationalizability and strong level-$k$ thinking is the assumption of uniform level-$1$ belief systems, this difference illustrates the impact of the initial beliefs assumption \emph{give}n the level of reasoning (i.e., level $3$) and the fact that both solution concepts feature the assumption that players are able to do forward induction.

The picture looks different for player 2. Only 43\% of player 2 choose $r$ conditional on the subgame is played. This is slightly less than for $k$-level strong rationalizability for $k \geq 2$ ($57\%$). Together this suggests that how participants view the context of the game, as captured by their initial beliefs, may depend on the player role. While the uniform beliefs assumptions seems largely consistent with the behavior of participants in the role of player 1, it is apparently not a descriptive assumption for the majority of participants in the role of player 2.

\begin{figure}[h!]\caption{Game used by Evdokimov and Rustichini (2016)\label{ER2016}}
	\centering
	\includegraphics[scale=0.4]{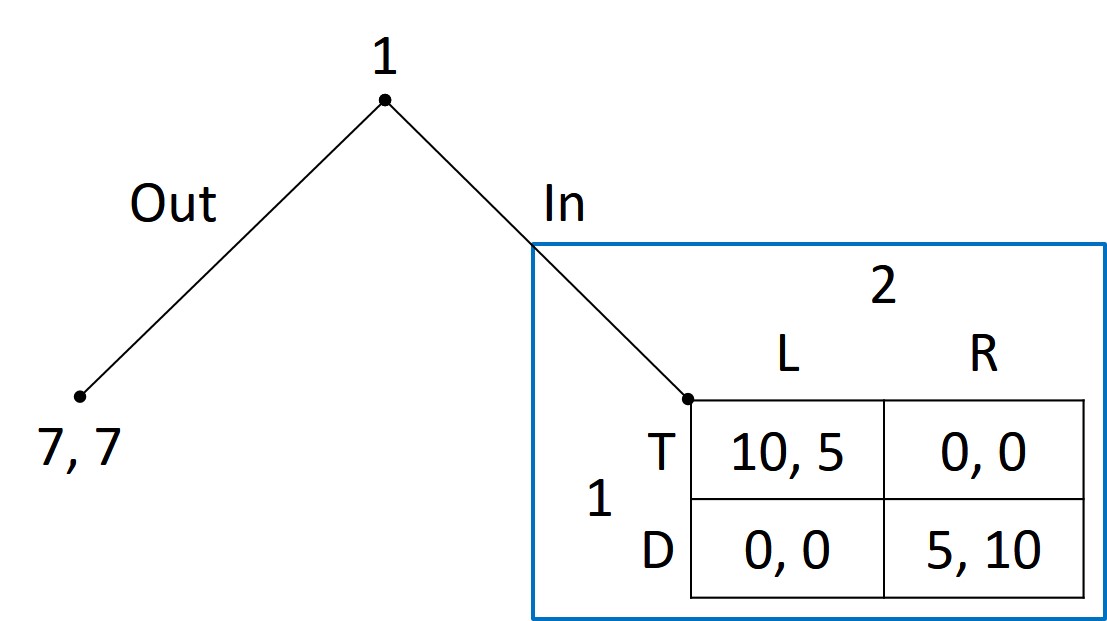}
\end{figure}

We can use an experiment by Evdokimov and Rustichini (2016) to check for the robustness of the last observation for player 2. Their experiment makes use of the Battle-of-sexes games with outside option depicted in Figure~\ref{ER2016}.
 This game has a similar best response structure as the Battle-of-the-sexes game with an outside option I (Example~\ref{BoSI_example}) and to the game used by Balkenborg and Nagel (2016). Thus, strategies consistent with various solution concepts are analogous to  Table~\ref{BN_table} (see Example~\ref{BoSI_example} for more detailed arguments).

There were 230 participants in the experiment. Participants played the game repeatedly. Their player roles could switch between repetitions. Between rounds, they received limited feedback: Player 1 received no feedback about player 2. Player 2 received information on whether or not the Battle-of-the-sexes subgame was reached, i.e., whether or not player 1 moved ``In''. There are different treatments that differ in the number of times the game had been repeated, when questions for belief elicitation were asked during the repetitions, and in the incentive structure.\footnote{Unfortunately, we do not have the belief elicitation data.} \begin{table}[h!]\caption{Choices in Evdokimov and Rustichini (2016)\label{ER_choices_table}}
\footnotesize
\centering
\begin{tabular}{||c||c|c||c|c||} \hline \hline & \multicolumn{2}{c||}{Strong level-$k$} & \multicolumn{2}{c||}{$k$-level extensive-form}  \\ & \multicolumn{2}{c||}{(with uniform level-$1$ belief system)} & \multicolumn{2}{c||}{rationalizability} \\ \hline \hline
Level & Player 1 & Player 2 & Player 1 & Player 2 \\ \hline \hline
$1$ & $62\%$ & $22\%$ & $98\%$ & $100\%$ \\
$2$ & $62\%$ & $22\%$ & $98\%$ & $78\%$ \\
$3$ & $62\%$ & $22\%$ & $36\%$ & $78\%$ \\ \hline \hline
\end{tabular}
\normalsize
\end{table}

Focusing on the behavioral data from all of their treatments, we classify individual choices as in Table~\ref{ER_choices_table}. Strong level-$k$ thinking with uniform level-$1$ belief systems predicts that player 1 always chooses Out at levels above 1, which is consistent with 62\% of the choices made by the participants. In terms of player 2's choice, the strong level-$k$ with uniform level-$1$ belief systems predicts that player 2 will select $R$, which is consistent with only 22\% of the choices. In contrast, the $k$-level strong rationalizability fits 78\% of the choices made by player 2. However, the strong rationalizability only fits 36\% of the player 1's choices. While the percentages differ from the corresponding percentages for the Balkenborg-Nagel game in Table~\ref{BN_choices_table}, the stylized fact from both experiments is that strong level-$k$ with uniform beliefs fits better to the behavior of player 1 while strong rationalizability fits better to player 2. Again, we conclude that while the uniform beliefs assumptions seems largely consistent with the behavior of participants in the role of player 1, it is not a descriptive assumption for the majority of participants in the role of player 2. By fitting both strong level-$k$ and level-$k$ strong rationalizability to the data, we can draw these conclusions about the role of uniform level-$1$ beliefs \emph{given} comparable levels of reasoning and the ability to do forward induction. We speculate that the uniform beliefs assumptions works for player 1 because the principle of insufficient reason is perhaps natural when a player has no prior experience with another player. In contrast, player 2 moves only after having observed some behavior of player 1, in which case it is not clear why he should still use the principle of insufficient reason. 








\section{Closing Remarks}

We extended normal-form level-$k$ thinking to games in extensive form by allowing for updating of beliefs during the play. Players can now use it these updated beliefs to make predictions over opponents' future play. In no way we want to suggest that strong level-$k$ thinking will be the ultimate behavioral solution concept that fits the data in games in extensive form better than other solution concepts. Quite to the contrary, we expect that in abstract choice environments some subjects in experiments may lack the ability to meaningfully draw conclusions from opponents' past play for predictions of opponents' future play. This ability is like a second dimension of sophistication that is distinct from (but interacts with) the binding cognitive bound. Our hope is that by applying strong level-$k$ thinking to experimental games and comparing it to normal-form level-$k$ thinking, we can learn about the prevalence of this second dimension of strategic sophistication.

We contrasted strong level-$k$ thinking with other existing iterative solutions concepts to game in extensive forms in order to emphasize that there is more than one way to approach levels of thinking in games in extensive form. Unfortunately, strong rationalizability and strong $\Delta$-rationalizability have been understudied in experimental game theory probably because there is no text-book treatment available of these solution concepts. However, similar to Kneeland (2015)'s demonstration that normal-form level-$k$ rationalizability is an interesting behavioral solution concept for the empirical study of levels of reasoning, we hope that smart experiments on level-$k$ strong rationalizability will emerge.  

The analysis of strong level-$k$ thinking could be taken further in both theoretical and experimental directions. For instance, it might be possible to extend the detailed epistemic analysis of the differences between normal-form level-$k$ thinking and $k$-level rationalizability by Brandenburger, Friedenberg, and Kneeland (2020) to strong level-$k$ thinking and $k$-level strong rationalizability. Moreover, while more existing data sets on experimental games in extensive form could be analyzed with strong level-$k$ thinking, we currently think about new experiments on games in extensive form tailor-made for testing strong level-$k$ thinking. 

We compared strong level-$k$ thinking with other level-$k$ thinking solution concepts to games such as normal-form level-$k$, $k$-level strong rationalizability/extensive-form rationalizability, $k$-level iterated admissibility, $k$-level backward rationalizability, backward level-$k$, and $k$-depth backward induction. Another useful comparison would be with the concurrent extension of the cognitive hierarchy model by Lin and Palfrey (2023). Another approach would extend dominance-$k$ to games in extensive form. Costa-Gomes, Crawford, and Broseta (2001) consider beside the level-$k$ model in games in normal form also a dominance-$k$ model defined as follows: There are $k$ rounds of elimination of actions dominated by a pure actions and a best response to a uniform belief over the remaining actions. This differs from $k$-level rationalizability in two respects. First, in its characterization of $k$-level rationalizability by $k$-iterated elimination of strictly dominated actions (Pearce, 1984), an action can also be eliminated when it is dominated by a \emph{mixed} action. Second, $k$-level rationalizability does not assume best response to a uniform belief over remaining actions but allows best responses to any belief over remaining actions. The dominance-$k$ model of Costa-Gomes, Crawford, and Broseta (2001) could also be extended to games in extensive form using ideas from iterated elimination of conditionally dominated strategies from Shimoji and Watson (1998). A strategy is conditionally dominated if it is dominated on the subspace of strategy profiles reaching an information set. With this idea, it is now straightforward to define the notion of conditional dominance-$k$ by first $k$-rounds elimination of strategies that are conditionally dominated by pure strategies and then best responses to uniform belief systems over remaining strategies at every information set. Yet another interesting approach could be based on forward induction in a backward inductive manner. Recently, Meier and Perea (2023) defined and characterized a version of rationalizability in extensive form featuring both forward and backward induction but giving priority to backward induction. It would possible to consider a version of it with fixed initial belief systems in the spirit of our notions of strong level-$k$ thinking and backward level-$k$ thinking.

\appendix

\section*{Proofs}

\subsection*{Proof of Proposition~\ref{relationship}}

We prove constructively using induction on the levels.

\noindent Base Case: For all $i \in N$, since $\beta^1_i \in B_i^1$, $L_i^1(\beta^1) \subseteq R_i^1$.

\noindent Inductive Hypothesis: For all $i \in N$ and $1 \leq \ell < k$, $L_i^{\ell}(\beta^1) \subseteq R_i^{\ell}$.

\noindent Inductive Step: We need to show that for every $i \in N$, $L_i^{k}(\beta^1) \subseteq R_i^{k}$. Pick any $a_i \in L_i^k(\beta^1)$. By Definition~\ref{levelk}, there exists a belief $\beta_i \in \Delta(A_{-i})$ that is certain of $L_{-i}^{k-1}(\beta^1)$ such that $a_i$ is rational with $\beta_i$. By the induction hypothesis, such a belief $\beta_i$ is also certain of $R_{-i}^{k-1}$. Hence, $a_i \in R_{-i}^k$. \hfill $\Box$

\subsection*{Proof of Proposition~\ref{converse}}

Since the game is finite, there exists $K$ such that for all $k \geq K$, $R_i^{\infty} = R_i^k$ for all $i \in N$.

For any $a_i \in R_i^{\infty}$ there exists $\beta_i \in \Delta(A_{-i})$ such that $\beta_i(R_{-i}^K) = 1$ and $a_i$ is rational for player $i$ given $\beta_i$. Since this holds for every player $i \in N$, set $\beta^1 = \beta = (\beta_i)_{i \in N}$. Then $a_i \in L^1_i(\beta^1)$. \hfill $\Box$

\subsection*{Proof of Proposition~\ref{EFL1_NFL1}}

For $k = 1$, note that for any player $i \in N$, if a strategy is rational with respect to a profile of full-support beliefs $\beta^1_i$ in the associated normal-form that is consistent with the profile of full-support belief systems $\bar{\beta}^1$, then it is rational with respect to $\bar{\beta}^1_i(I_i)$ conditional on reaching information set $I_i$. This is due to the fact that conditioning does not alter the relative likelihoods of opponent's strategies reaching the information set. Conversely, if a strategy is rational with respect to a profile of system of full-support beliefs $\bar{\beta}^1$ at \emph{every} information set $I_i \in \mathcal{I}_i$, then is also rational with respect to the profile of full-support beliefs $\beta^1$ in the associated normal-form that is consistent with $\bar{\beta}^1$.\hfill $\Box$

\subsection*{Proof of Proposition~\ref{refinement}}

We prove by induction on the levels.

\noindent Base Case: By Proposition~\ref{EFL1_NFL1}, $\bar{L}_i^1(\bar{\beta}^1) = L_i^1(\beta^1)$ for all $i \in N$. Thus, $\bar{L}^1(\bar{\beta}^1) = L^1(\beta^1)$ and $Z(\bar{L}^1(\bar{\beta}^1)) = Z(L^1(\beta^1))$.

\noindent Inductive Hypothesis: For any $\ell$ with $1 \leq \ell < k$, $Z(\bar{L}^{\ell}(\bar{\beta}^1)) \subseteq Z(L^{\ell}(\beta^1))$.

\noindent Inductive Step: We need to show $Z(\bar{L}^{k}(\bar{\beta}^1)) \subseteq Z(L^{k}(\beta^1))$. Let $z \in Z(\bar{L}^{k}(\bar{\beta}^1))$. 

Let $N(z) \subseteq N$ be the set of players $i \in N$ for whom an information set $I_i$ is reached along the path to $z$. Since the game in extensive form has perfect recall, each player's set of information sets form an arborescence, i.e., there is a partial order on the set of information sets that orders information sets by precedence. Since it is a partial order, it may have upper bounds, i.e., initial information sets. Yet, along each path, there is a unique upper bound. For any $i \in N(z)$, let $I_i$ denote this first information set of $i$ along the path to $z$.

Since $z \in Z(\bar{L}^{k}(\bar{\beta}^1))$, there exists $s \in \bar{L}^{k}(\bar{\beta}^1)$ with $z(s) = z$. For any $i \in N(z(s))$, there exists $\bar{\beta}_i \in \bar{B}_i^k(\bar{\beta}^1)$ such that $\bar{\beta}_i(I_i)(\bar{L}_{-i}^{k-1}(\bar{\beta}^1)) = 1$ and $s_i$ is rational at $I_i$ with $\bar{\beta}_i(I_i)$ (with $s_i$ being the $i$-component of strategy profile $s$). 

For any nonempty $Z' \subseteq Z$, with some slight abuse of notation denote by $S_{-i}(Z') = \{s_{-i} \in S_{-i} : z(s_i, s_{-i}) \in Z', s_i \in S_i\}$. Note that for any nonempty $Z', Z'' \subseteq Z$ with $Z' \subseteq Z''$ implies $S_{-i}(Z') \subseteq S_{-i}(Z'')$. The induction hypothesis, $Z(\bar{L}^{k-1}(\bar{\beta}^1)) \subseteq Z(L^{k-1}(\beta^1))$, implies $Z(S_i \times \bar{L}_{-i}^{k-1}(\bar{\beta}^1)) \subseteq Z(S_i \times L_{-i}^{k-1}(\beta^1))$. Hence, we have $S_{-i}(Z(S_i \times \bar{L}_{-i}^{k-1}(\bar{\beta}^1))) \subseteq S_{-i}(Z(S_i \times L_{-i}^{k-1}(\beta^1)))$.

Note that for any nonempty $S'_{-i} \subseteq S_{-i}$, $S_{-i}(Z(S_i \times S'_{-i})) \supseteq S'_{-i}$. Thus, $\bar{\beta}_i(I_i)(\bar{L}_{-i}^{k-1}(\bar{\beta}^1)) = 1$ implies $\bar{\beta}_i(I_i)(S_{-i}(Z(S_i \times \bar{L}_{-i}^{k-1}(\bar{\beta}^1)))) = 1$ and $\bar{\beta}_i(I_i)(S_{-i}(Z(S_i \times L_{-i}^{k-1}(\beta^1)))) = 1$.

Define $\beta_i = \bar{\beta}_i(I_i)$. Note that opponents strategies in $S_{-i}(Z(S_i \times L_{-i}^{k-1}(\beta^1))) \setminus L_{-i}^{k-1}(\beta^1)$ do not affect terminal histories, i.e., $Z(S_i \times S_{-i}(Z(S_i \times L_{-i}^{k-1}(\beta^1)))) \subseteq Z(S_i \times L_{-i}^{k-1}(\beta^1))$. To see this, consider any $z(\tilde{s}_1, \tilde{s}_{-i}) \in Z(S_i \times S_{-i}(Z(S_i \times L_{-i}^{k-1}(\beta^1))))$ with $\tilde{s}_i \in S_i$ and $\tilde{s}_{-i} \in S_{-i}(Z(S_i \times L_{-i}^{k-1}(\beta^1)))$. Now, since $\tilde{s}_{-i} \in S_{-i}(Z(S_i \times L_{-i}^{k-1}(\beta^1)))$, there exists $\hat{s}_{i} \in S_{i}$ such that $z(\hat{s}_i, \tilde{s}_{-i}) \in Z(S_i \times L_{-i}^{k-1}(\beta^1))$. Because this holds for any $\hat{s}_i \in S_i$, we can set $\hat{s}_i = \tilde{s}_i$. Then $z(\tilde{s}_i, \tilde{s}_{-i}) \in Z(S_i \in L_{-i}^{k-1}(\beta^1))$, which is exactly what we needed to show.

We conclude that $\beta_i( \cdot \mid  L_{-i}^{k-1}(\beta^1))$ yields the same expected utilities from strategies as $\beta_i$, where $\beta_i( \cdot \mid  L_{-i}^{k-1}(\beta^1))$ is $\beta_i$ conditional on $L_{-i}^{k-1}(\beta^1)$. Moreover, $\beta_i( \cdot \mid  L_{-i}^{k-1}(\beta^1)) \in B_{i}^{k}(\beta^1)$.

Since $s_i$ is rational with $\bar{\beta}_i(I_i)$ at $I_i$, it is also rational with $\beta_i( \cdot \mid  L_{-i}^{k-1}(\beta^1))$. Thus, $s_i \in L_i^k(\beta^1)$.

This holds for all $i \in N(z(s))$. Since strategies of any other player $j \in N \setminus N(z(s))$ do not affect reaching $z(s)$, we can choose any $s'_j \in L_j^k(\beta^1)$. Note that $z\left((s_i)_{i \in N(z(s))}, (s'_j)_{j \in N \setminus N(z(s))}\right) = z(s)$. Since $z\left((s_i)_{i \in N(z(s))}, (s'_j)_{j \in N \setminus N(z(s))}\right) \in Z(L^k(\beta^1))$, we have $z(s) \in Z(L^k(\beta^1))$.  \hfill $\Box$

\subsection*{Proof of Remark~\ref{first-level_equivalence}} 

For every player $i \in I$, we have that strategy $s_i \in \vec{L}_i^1(\bar{\beta}^1)$ if and only if $s_{i \mid I_i}$ is rational at $I_i$ given $\bar{\beta}_i^1(I_i)$ for every $I_i \in \mathcal{I}_i$. This is implies that $s_i$ is rational at $I_i$ given $\bar{\beta}_i^1(I_i)$ for every $I_i \in \mathcal{I}_i$. Thus, $s_i \in L_i^1(\bar{\beta}^1)$. The converse does not hold necessarily since we can have the case that $s_i$ does not reach $I_i$. In such a case $s_i$ is trivially rational at $I_i$ but $s_{i \mid I_i}$ may not be rational at $I_i$. However, fix an backward level-$1$ outcome. Consider the path of information sets towards this outcome. Since we assumed that information sets in the game in extensive-form are ordered by a precedence relation, this is well-defined. Consider a profile of strategies which allows the outcome to be reached. Then they also allow to reach any information set on the path towards the outcome. Along any such information set $I_i$ we have that $s_i$ is rational at $I_i$ given $\bar{\beta}_i^1(I_i)$ if and only if $s_{i \mid I_i}$ is rational at $I_i$ given $\bar{\beta}_i^1(I_i)$.\hfill $\Box$ \\

\end{document}